\documentclass[letter,superscriptaddress,twocolumn,prb,showkeys,showpacs]{revtex4-1}

\usepackage[utf8]{inputenc}
\usepackage{amsmath}
\usepackage{amssymb}
\usepackage{amsfonts}     
\usepackage{graphicx}   
\usepackage{verbatim}  
\usepackage{color}     
\usepackage{subfigure}  
\usepackage{hyperref}   
\usepackage{float}
\usepackage{natbib}
\usepackage{gensymb}
\usepackage{color}
\usepackage{epstopdf}
\usepackage{multirow}

\begin{document}

\title{Examining the surface phase diagram of IrTe$_2$ with photoemission.}

\author{M.~Rumo}
\altaffiliation{Corresponding author:\\ maxime.rumo@unifr.ch}
\affiliation{D{\'e}partement de Physique and Fribourg Center for Nanomaterials, Universit{\'e} de Fribourg, CH-1700 Fribourg, Switzerland}

\author{C.W.~Nicholson}
\affiliation{D{\'e}partement de Physique and Fribourg Center for Nanomaterials, Universit{\'e} de Fribourg, CH-1700 Fribourg, Switzerland}

\author{A.~Pulkkinen}
\affiliation{D{\'e}partement de Physique and Fribourg Center for Nanomaterials, Universit{\'e} de Fribourg, CH-1700 Fribourg, Switzerland}
\affiliation{School of Engineering Science, LUT University, FI-53850, Lappeenranta, Finland}

\author{B.~Hildebrand}
\affiliation{D{\'e}partement de Physique and Fribourg Center for Nanomaterials, Universit{\'e} de Fribourg, CH-1700 Fribourg, Switzerland}

\author{G.~Kremer}
\affiliation{D{\'e}partement de Physique and Fribourg Center for Nanomaterials, Universit{\'e} de Fribourg, CH-1700 Fribourg, Switzerland}

\author{B.~Salzmann}
\affiliation{D{\'e}partement de Physique and Fribourg Center for Nanomaterials, Universit{\'e} de Fribourg, CH-1700 Fribourg, Switzerland}

\author{M.-L.~Mottas}
\affiliation{D{\'e}partement de Physique and Fribourg Center for Nanomaterials, Universit{\'e} de Fribourg, CH-1700 Fribourg, Switzerland}

\author{K.Y.~Ma}
\affiliation{Department of Chemistry, University of Zurich, CH-8000 Zurich, Switzerland}

\author{E~L.~Wong}
\affiliation{Femtosecond Spectroscopy Unit, Okinawa Institute of Science and Technology Graduate University, 1919-1 Tancha, Onna-son, Kunigami, Okinawa 904-495, Japan}

\author{M.K.L.~Man}
\affiliation{Femtosecond Spectroscopy Unit, Okinawa Institute of Science and Technology Graduate University, 1919-1 Tancha, Onna-son, Kunigami, Okinawa 904-495, Japan}

\author{K.~M.~Dani}
\affiliation{Femtosecond Spectroscopy Unit, Okinawa Institute of Science and Technology Graduate University, 1919-1 Tancha, Onna-son, Kunigami, Okinawa 904-495, Japan}

\author{B.~Barbiellini}
\affiliation{School of Engineering Science, LUT University, FI-53850, Lappeenranta, Finland}
\affiliation{Department of Physics, Northeastern University, Boston, Massachusetts 02115, USA}

\author{M.~Muntwiler}
\affiliation{Paul Scherrer Institute, CH-5232 Villigen PSI, Switzerland}

\author{T.~Jaouen}
\affiliation{D{\'e}partement de Physique and Fribourg Center for Nanomaterials, Universit{\'e} de Fribourg, CH-1700 Fribourg, Switzerland}

\author{F.~O.~von~Rohr}
\affiliation{Department of Chemistry, University of Zurich, CH-8057 Zurich, Switzerland}

\author{C.~Monney}
\altaffiliation{Corresponding author:\\ claude.monney@unifr.ch}
\affiliation{D{\'e}partement de Physique and Fribourg Center for Nanomaterials, Universit{\'e} de Fribourg, CH-1700 Fribourg, Switzerland}

\begin{abstract}
In the transition metal dichalcogenide IrTe$_2$, low-temperature charge-ordered phase transitions involving Ir dimers lead to the occurrence of stripe phases of different periodicities, and nearly degenerate energies. Bulk-sensitive measurements have shown that, upon cooling, IrTe$_2$ undergoes two such first-order transitions to $(5\times1\times5)$ and $(8\times1\times8)$ reconstructed phases at $T_{c_1}\sim 280$~K and $T_{c_2}\sim 180$~K, respectively. Here, using surface sensitive probes of the electronic structure of IrTe$_2$, we reveal the first-order phase transition at $T_{c_3}=165$~K to the $(6\times1)$ stripes phase, previously proposed to be the surface ground state. This is achieved by combining x-ray photoemission spectroscopy and angle-resolved photoemission spectroscopy, which give access to the evolution of stripe domains and a particular surface state, the energy of which is dependent on the Ir dimer length. By performing measurements over a full thermal cycle, we also report the complete hysteresis of all these phases.
\end{abstract}
\date{June 4, 2020}
\maketitle

\section{Introduction}

Transition-metal dichalcogenides (TMDCs) are layered quasi-two dimensional (2D) materials that have generated considerable interest in recent years due to the possibility of reducing their thickness down to the monolayer as well as to their particularly diverse optical and electronic properties despite their chemical simplicity~\cite{WangNat,RadisavljevicNat,MakNat,BertoniPRL}. Additionally, TMDCs have been extensively studied for several decades, due to the occurrence of phase transitions such as charge-density waves (CDWs) or superconductivity~\cite{RossnagelIOP,JohannesPRB,PyonJPSJ} at low temperatures. An open question is how these collective states evolve for thicknesses of a few layers at surfaces. Many recent examples have illustrated different behaviors in monolayers, namely an enhanced critical temperature for the CDW in TiSe$_2$~\cite{ChenNanoLet}, enhanced superconductivity in TaS$_2$~\cite{NavarroNatCom}, or a change in the symmetry of the CDW in VSe$_2$~\cite{ChenPRL}. In this context, the surface of IrTe$_2$ offers an exciting platform for studying ordered phases in a quasi-2D material with large spin-orbit coupling on the transition metal site. A complex succession of charge-ordered phases involving the creation of Ir dimers~\cite{PascutPRL,MauererPRB,HsuPRL} has been observed in IrTe$_2$ at low temperature, which gives way to superconductivity for thin samples~\cite{YoshidaSC}, after rapid cooling~\cite{OikeSuper} or with Pt substitution~\cite{PyonJPSJ}.
\\

IrTe$_2$ undergoes a first-order structural phase transition at $T_{c_1}\sim280$~K from a trigonal CdI$_2$-type ($P\overline{3}m1$) unit cell to a monoclinic ($P\overline{1}$) unit cell accompanied by jumps in the resistivity and magnetic susceptibility~\cite{KoNatCom,FangScienRep,JobicZeit,MatsumotoJLTP,ToriyamaJapanLetters,KoleySSCom,LiSciRep,ParisXAS}. In this first low-temperature charge-ordered phase, one-dimensional stripes of Ir dimers with a strongly reduced bond length have been observed by x-ray diffraction and are described by a wave vector $(5\times1\times5)$~\cite{PascutPRL,PascutPRB,KoNatCom,LiNatCom,ToriyamaJapanLetters,OhPRL,TakuboPRB}. At $T_{c_2}\sim180$~K, a second phase transition follows and the charge-ordering wave vector of this new low-temperature phase is $(8\times1\times8)$ in the bulk of IrTe$_2$. This has stimulated many scanning tunneling microscopy (STM) studies, which evidenced additional ordering patterns and revealed a surface $(6\times1)$ periodicity proposed to be the ground state reconstruction~\cite{MauererPRB,KoNatCom,LiNatCom,DaiPRB}. In addition, a detailed low-energy electron diffraction (LEED) and STM study observed in addition the coexistence of the surface $(8\times1)$ and $(6\times1)$ phases over a wide temperature range~\cite{ChenPhysRev}. It was later realized that the $(6\times1\times6)$ phase can also appear in the bulk at very low temperature, if the cooling rate is higher than 4 K$/$min~\cite{TakuboXRD}. Numerous angle-resolved photoelectron spectroscopy (ARPES) investigations have exposed large changes in the spectral weight of electronic states up to $2$~eV below the Fermi level ($E_F$) at low temperatures, leading to severe band broadening~\cite{MauererPRB,ChenPhysRev,KoNatCom,LiNatCom,DaiPRB,MonneyPRB}. In parallel, x-ray photoemission spectroscopy (XPS) studies have identified a large splitting of the Ir $4f$ core-levels across the charge-ordered phase transitions into Ir$^{+3}$ and Ir$^{+4}$ mixed valence states~\cite{KoNatCom,QianIOP}. The charge-ordered phases are stabilized by the energy gain due to the stronger bonds of the dimerized states, despite the elastic energy loss \cite{PascutPRB}. Therefore, phases of different stripe periodicities appear with increasing dimer densities as temperature decreases. Their complex kinetics, involving dimers breaking and reassembling, has been studied by Mauerer \textit{et al.} \cite{MauererPRB}. Interestingly, a very recent study predicts the correct $T_{c_1}$ by considering the entropy of the $(5\times1\times5)$ phase~\cite{SalehEntropy}. However, in all experimental studies, a clear sequence of sharp transitions between the surface $(5\times1)$, $(8\times1)$ and $(6\times1)$ phases could not be unambiguously detected.
\\

In this paper, we present a systematic temperature-dependent study of the electronic properties at the surface of IrTe$_2$. Using XPS, we measure the evolution of the Ir dimer density upon cooling down to $30$~K and warming back to room temperature by quantifying the Ir $4f$ core level peak intensities. While we confirm the sharp first-order transition occurring at $T_{c_1}$, we observe a more intricate behavior below $T_{c_2}$ at the surface. The $(5\times1)$ phase is replaced by the $(8\times1)$ phase that progressively changes into the $(6\times1)$ phase, indicating $(6\times1)$ domain growth at the expense of the $(8\times1)$ domains. However, our ARPES measurements reveal that a \textit{third} first-order structural transition between the $(8\times1)$ and $(6\times1)$ phases occurs at $T_{c_3}=165$~K. This is based on the observation of a surface state at about $1$~eV binding energy that is an excellent marker of the periodicity of the dominating phase, since its binding energy is dictated by the Ir dimer length, as supported by density functional theory (DFT) calculations. By analyzing both XPS and ARPES data over the full warming and cooling cycle, and combining them with LEED measurements, we are able to reconstruct the complete surface phase diagram of IrTe$_2$ with the hysteretic behavior of all $(5\times1)$, $(8\times1)$, and $(6\times1)$ phases.
\\

\section{Methods}

Single crystals of IrTe$_2$ were grown using the self-flux method~\cite{FangScienRep,JobicZeit}. They were characterized by magnetic susceptibility and resistivity measurements (see Appendix \ref{sec_App_transport}), which confirm that $T_{c_1}$ $= 278$~K and $T_{c_2}$ = $180$~K. Samples were cleaved at room temperature in vacuum at a pressure of about $10^{-8}$~mbar; during the photoemission measurements, the base pressure was better than $5\times10^{-11}$~mbar. XPS measurements were acquired at the PEARL beamline~\cite{PEARL} of the Swiss Light Source. The total energy resolution was $190$~meV. The temperature-dependent ARPES study was carried out using a Scienta DA$30$ photoelectron analyzer and monochromatized He$_I$ radiation as excitation source ($h\nu=21.22$~eV). The total energy resolution was about $5$~meV and the error on the sample temperature was estimated to be $5$~K. Cooling and warming of the sample were carried out at rates $<8$ K$/$min and $\sim2$ K$/$min, respectively, and each  measurement was preceded by a pause of at least 10 min, to ensure thermalization. Small spot LEED with micron resolution ($\mu$LEED) data were obtained using a low energy electron microscopy (LEEM)/photoemission electron microscopy (PEEM)/LEED instrument (Elmitec GmbH). Samples were cleaved in ultra-high vacuum ($1\times 10^{-10}$~mbar). The aperture for LEED measurements was set to select a spatial region on the sample of 20~$\mu$m. 

DFT calculations with spin-orbit interaction were performed using the \textsc{Vienna ab-initio simulation package} (VASP)~\cite{kresse1993,kresse1994,kresse1996a,kresse1996b} within the projector augmented wave method~\cite{kresse1999} and the Perdew-Burke-Ernzerhof (PBE) functional~\cite{perdew1996}. The cutoff energy was set to $400$~eV and the k-point grid spacing was $<0.02$~\AA$^{-1}$. Band unfolding has been performed using the \textsc{BandUP} code~\cite{medeiros2014,medeiros2015}.
\\

\section{Results and discussion}

\subsection{Ir $4f$ core levels}

\begin{figure}[t]
\includegraphics[width=0.9\columnwidth]{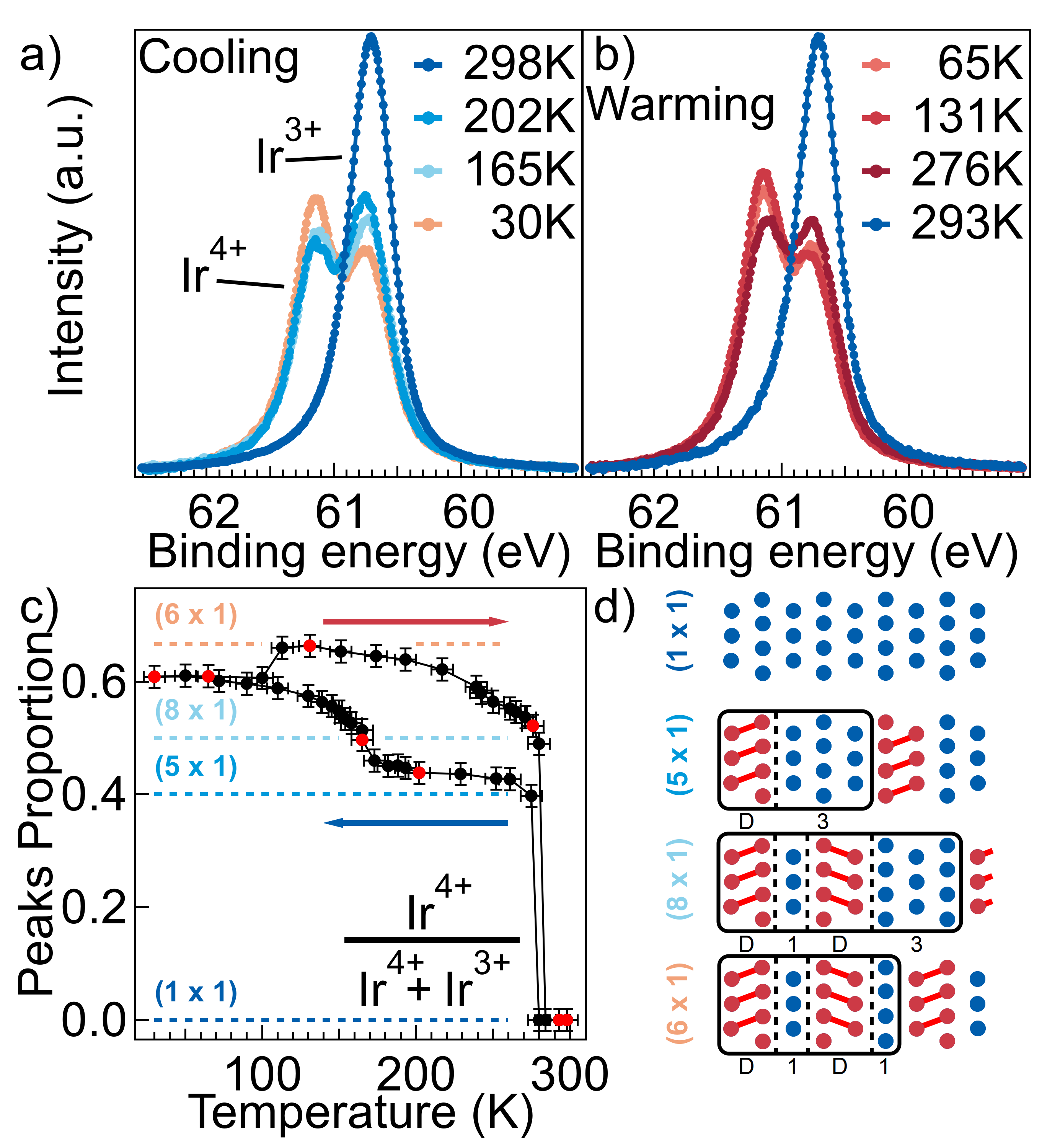}
\caption{\label{Figure XPS} 
(Color online)~XPS spectra of Ir $4f_{7/2}$ core levels measured with a photon energy $h\nu =200$~eV at various temperatures during (a)~cooling and (b)~warming. (c)~Intensity ratio of the Ir$^{4+}$/(Ir$^{3+}$ + Ir$^{4+}$) peaks in the Ir $4f$ core levels as a function of temperature. (d)~Schematic description of the stripe periodicities in the Ir planes for different phases.
} 
\end{figure}

We present first a detailed XPS study of the Ir $4f$ core levels. Figures~\ref{Figure XPS}~(a) and \ref{Figure XPS}~(b) shows a zoom on the Ir $4f_{7/2}$ core-levels measured at different temperatures upon cooling and warming, respectively. A clear splitting occurs below $T_{c_1}$, with a new peak appearing at $61.2$~eV binding energy and corresponding to the Ir$^{4+}$ states (the peak at $60.6$~eV binding energy is attributed to the Ir$^{3+}$ states). Across $T_{c_2}$ upon cooling [ Fig.~\ref{Figure XPS}~(a) ], the intensity ratio between the Ir$^{4+}$ and Ir$^{3+}$ peaks is reversed and, upon warming [Fig.~\ref{Figure XPS}~(b)], this intensity ratio changes further in a non-trivial way. In Fig.~\ref{Figure XPS}~(c), the intensity ratio (area) Ir$^{4+}$/(Ir$^{3+}$+Ir$^{4+}$) is plotted as a function of temperature, when cooling from $298$~K down to $30$~K and then warming back to $298$~K.  

\begin{figure*}
\includegraphics[width=1.8\columnwidth]{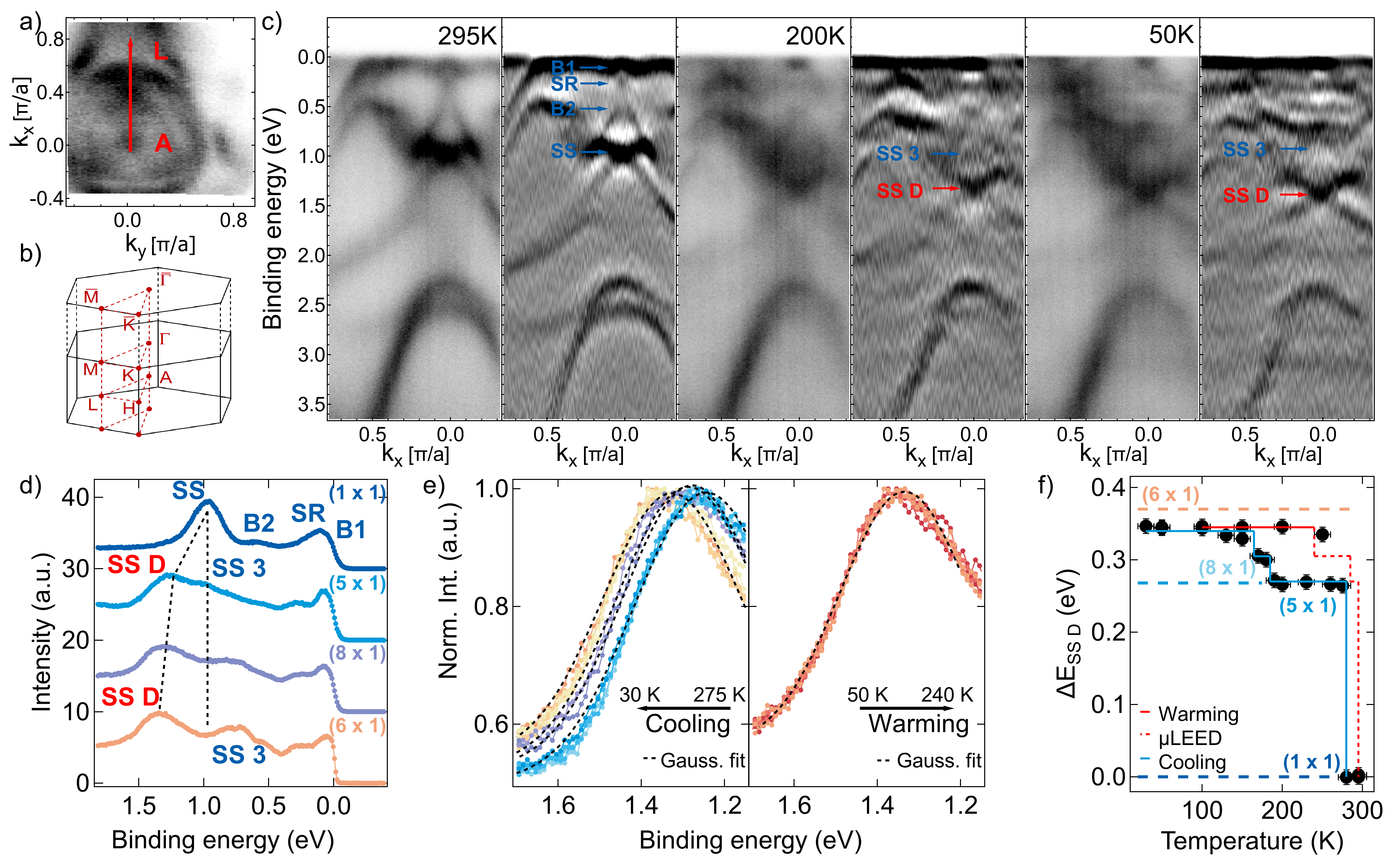}
\caption{\label{Figure UV} 
(Color online) (a)~Fermi surface mapping of IrTe$_2$ for $h\nu = 21.22$~eV taken at $295$~K. (b)~IrTe$_2$ Brillouin zone. (c)~ARPES spectra measured along AL direction for $h\nu = 21.22$~eV at three different temperatures with their respective second derivatives. (d)~Representative EDCs for the $(1\times1)$ (at $295$~K), $(5\times1)$ (at $200$~K), $(8\times1)$ (at $165$~K) and $(6\times1)$ (at $50$~K) phase (integrated $\pm0.06$~\AA$^{-1}$ around A along AL direction). (e)~Normalized temperature-dependent EDCs and their respective fits (Gaussian function) upon cooling and warming. (f)~Binding energy of the surface state SS D measured in ARPES as a function of temperature (full symbols) and extracted from DFT (thick dashed lines), together with the transition temperature obtained from $\mu$LEED (red dashed line).
} 
\end{figure*}

The Ir$^{4+}$/(Ir$^{3+}$+Ir$^{4+}$) intensity ratio measured by XPS has been interpreted as a measure of the density of Ir$^{4+}$ $-$Ir$^{4+}$ dimers in the different phases observed in IrTe$_2$~\cite{KoNatCom,QianIOP}. On cooling, different stripe periodicities have been evidenced by STM and LEED at the surface of IrTe$_2$~\cite{HsuPRL,ChenPhysRev,MauererPRB,KoNatCom}. Below $T_{c_1}$, a $(5\times1)$ phase with two dimerized Ir$^{4+}$ atoms, labeled \textbf{D}, and triple undimerized Ir atoms have been observed [Fig.~\ref{Figure XPS}~(d)], giving a Ir$^{4+}$/(Ir$^{3+}$+Ir$^{4+}$) ratio of $0.4$ in good agreement with our XPS data just below $T_{c_1}$ [Fig.~\ref{Figure XPS}~(c)]. Across $T_{c_2}$, the $(5\times1)$ phase gives way to a $(8\times1)$ phase with a Ir$^{4+}$/(Ir$^{3+}$+Ir$^{4+}$) ratio of $0.5$, due to the presence of Ir$^{4+}$ dimers alternating with triple and single undimerized Ir atoms [Fig.~\ref{Figure XPS}~(d)]. At even lower temperatures, recent STM and LEED studies~\cite{HsuPRL,ChenPhysRev,MauererPRB,LiNatCom}, supported by DFT calculations, revealed a $(6\times1)$ phase with a Ir$^{4+}$/(Ir$^{3+}$+Ir$^{4+}$) ratio of $0.66$ and concluded that it represents the low-temperature ground state at the surface of IrTe$_2$. Indeed, below $T_{c_2}$, the Ir$^{4+}$/(Ir$^{3+}$+Ir$^{4+}$) ratio increases progressively above $0.45$ [Fig.~\ref{Figure XPS}~(c)], meaning that the $(8\times1)$ phase is gradually replaced by the $(6\times1)$ phase. At $30$~K, the $(6\times1)$ phase dominates over the $(8\times1)$ phase, since the Ir$^{4+}$/(Ir$^{3+}$+Ir$^{4+}$) ratio reaches a value of $0.61$, close to the maximum value of $0.66$ expected for a pure $(6\times1)$ phase with two dimers over six Ir atoms [see Fig.~\ref{Figure XPS}~(d)]. Interestingly, when warming the sample above $110$~K, this ratio increases to $0.65$, indicating further changes in the $(6\times1)$ vs $(8\times1)$ phase ratio. This effect, although small, can be directly seen on the XPS spectra of Fig.~\ref{Figure XPS}~(b), and will be addressed further below. In summary, upon cooling, the Ir$^{4+}$/(Ir$^{3+}$+Ir$^{4+}$) intensity ratio measured by XPS reveals that a sharp transition occurs across $T_{c_1}$, but that the evolution below $T_{c_2}$ is continuous with temperature, as a consequence of the coexistence of domains with different stripe periodicities.

\begin{figure*}
\includegraphics[width=2\columnwidth]{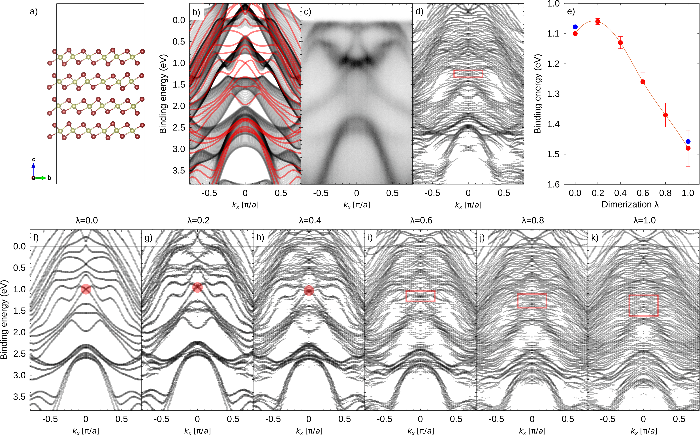}
\caption{(a) Four-layer slab model of the $(6\times1)$ phase surface. (b) DFT bulk projected bands and band structure of a four-layer slab along $\bar{\Gamma}\bar{M}$ in the $(1\times1)$ phase. (c) ARPES spectrum of IrTe$_2$ measured along AL direction for $h\nu = 21.22$~eV at 295~K. (d) DFT band structure for a four-layer slab along $\bar{\Gamma}\bar{M}$ in the $(5\times1)$ phase. (e) Binding energy of the surface state in interpolated structures as a function of the dimerization parameter $\lambda$. The solid blue circles represent the experimental binding energy of the surface state in the $(1\times1)$ and $(6\times1)$ phases. (f) - (k) DFT band structure for a four-layer slab along $\bar{\Gamma}\bar{M}$ in the $(6\times1)$ phase for $\lambda = 0.0 - 1.0$ (see text).\label{Figure DFT}} 
\end{figure*}

\subsection{Low-energy electronic structure}

We have also performed ARPES measurements as a function of temperature, to discriminate further the occurrence of difference phases in IrTe$_2$. Figure~\ref{Figure UV}~(a) shows its room-temperature Fermi surface (integrated over $0.05$~eV around $E_F$). At this photon energy, states close to the ALH plane are probed, in agreement with the literature~\cite{OotsukiJapanLettersES,OotsukiJapanLettersPt,OotsukiJournPhys}. The three-dimensional Brillouin zone and its surface projection are presented in Fig.~\ref{Figure UV}~(b). In Fig.~\ref{Figure UV}~(c), ARPES spectra taken at $295$~K~$>$~$T_{c_1}$, $T_{c_1} > 200$~K $ > $ $T_{c_2}$ and $50$~K~$ < $ $T_{c_2}$, along the AL direction are displayed, together with their second derivative. Corresponding energy distribution curves (EDCs) integrated around A are shown in Fig.~\ref{Figure UV}~(d). At $295$~K [Fig.~\ref{Figure UV}~(c), left panels], the electronic bands are sharp and, by comparison with the literature~\cite{LeeIOP,KingNatMat}, we can identify the presence of a bulk state B1 just below $E_F$, a surface resonance SR dispersing around $0.5$~eV binding energy on top of a bulk band B2, and an intense surface state SS at about $1$~eV binding energy. This is further confirmed by a DFT calculation of the bulk-projected band structure combined with a slab calculation [see Fig.~\ref{Figure DFT}~(b) and next paragraph]. All these features can be seen in the corresponding EDC [Fig.~\ref{Figure UV}~(d)]. When decreasing the temperature below $T_{c_1}$, the electronic bands become intricate due to the new translational symmetry of the charge-ordered phases and their mixed orientations~\cite{HsuPRL}. One distinguishes a multitude of folded bands [see Fig.~\ref{Figure UV}~(c) center and right panels], especially, in the binding energy range between $E_F$ and $2.0$~eV. At $200$~K, in the $(5\times1)$ phase, the surface state is split into two states. This is more obvious in the EDC (light blue curve) in Fig.~\ref{Figure UV}~(d). We attribute the surface state positioned at $1$~eV binding energy (labelled SS 3) as originating from the undimerized triple Ir atoms, since it lies at the same energy as the surface state of the $(1\times1)$ phase, for which there are only undimerized atoms. The intense second surface state (labelled SS D) is shifted to about $1.3$~eV binding energy and we attribute it to the dimerized Ir atoms. At $50$~K, in the $(6\times1)$-dominated phase [see right panels in Fig.~\ref{Figure UV}~(c) and orange EDC in Fig.~\ref{Figure UV}~(d)], the SS D surface state shifts further to higher binding energy. Looking at the ARPES spectra [graphs (c)], one sees that the surface state dispersion remains mostly unchanged across the phase transitions, except for the global energy shift, confirming its persistence at low temperature.

\subsection{DFT analysis and the surface-state}

To support our interpretation of the evolution of the binding energy of the surface state across the phase transitions, we have performed DFT calculations using a slab geometry for the $(1\times1)$, $(5\times1)$, and $(6\times1)$ surface phases, based on the atomic bulk structure of the corresponding phases~\cite{JobicZeit,PascutPRB} [see Fig.~\ref{Figure DFT}~(a) for the atomic structure of this slab in the $(6\times1)$ phase]. We have assumed that the structural parameters of the bulk IrTe$_2$ $(6\times1)$ phase are the same as in IrTe$_{2-x}$Se$_x$ $(x = 0.4)$~\cite{PascutPRB}. In the room-temperature phase, the surface state is identified by comparing the bulk projected band structure to the four-layer $(1\times1)$ slab band structure [Fig.~\ref{Figure DFT}~(b)]. The band crossing at $\bar{\Gamma}$ ($k=0$) near $1$~eV binding energy in the slab band structure appears in a gap region of the bulk projected band structure, thus confirming the surface nature of this state, in agreement with literature~\cite{LeeIOP,KingNatMat}. The binding energy of the surface state is well reproduced in comparison to the experiment at 295~K [Fig.~\ref{Figure DFT}~(c)]. In Fig.~\ref{Figure DFT}~(d), we show the unfolded four-layer slab band structure for the $(5\times1)$ phase. Here the band structure becomes complicated because of the multiplicity of the bands. A direct comparison with the four-layer $(1\times1)$ slab band structure [Fig.~\ref{Figure DFT}~(b)] helps to locate the surface state crossing at $k=0$ at about $1.25$~eV binding energy. The red box in Fig.~\ref{Figure DFT}~(d) renders the uncertainty about the exact localization of this surface state, which is likely duplicated due to the new surface periodicity.

Next we use DFT calculations to obtain the binding energy of this surface state in the $(6\times1)$ phase. Again, the band structure becomes complicated, because of the periodicity of the phase. In that respect, it is also important to recall that the dimer bond length $L_{\text{dim}}$ displays a significant variation as a function of the phase periodicity. Pascut \textit{et al.} inferred $L_{\text{dim}} = 3.119$~\AA\ and $3.099$~\AA\ for the $(5\times1)$ and for the $(8\times1)$ bulk phases, respectively, and proposed a value of $L_{\text{dim}} = 3.005$~\AA\ (confirmed experimentally in the STM study of Hsu \textit{et al.}~\cite{HsuPRL}) for a DFT-calculated $(6\times1)$ reconstruction~\cite{PascutPRB}. The overall picture is that the Ir dimer bond length decreases across the different phases. So it is very likely that the increase of the binding energy of the surface state SS D is due to a shortening of the Ir dimer bonds. To check this idea, we introduce an interpolation parameter $\lambda$ in our four-layer slab structural model for the surface $(6\times1)$ phase, with values $\lambda = 0.0$ and $\lambda = 1.0$ corresponding to the $(1\times1)$ surface phase and the surface $(6\times1)$ phase, respectively. To have a controlled interpolation of the structural parameters (lattice parameters and atom positions, as well as dimer length), structural relaxation was not considered. For comparison with experimental ARPES data, the surface $(6\times1)$ band structures were unfolded to a corresponding $(1\times1)$ reciprocal cell.
Figures~\ref{Figure DFT}~(f)$-$\ref{Figure DFT}~(k) show the corresponding DFT band structures for $\lambda = 0.0$ to $1.0$ along the $\bar{\Gamma}\bar{M}$ direction. In comparison to the calculations of Fig.~\ref{Figure DFT}~(b) for the $(1\times1)$ phase, the bands become back-folded to the original Brillouin zone with various weights. However, the surface state is still recognizable and shifts to higher binding energy. We have performed such calculations for $\lambda$ up to $1$ and tracked the energy position of the surface state. The resulting values for the surface state, shown in Fig.~\ref{Figure DFT}~(e) (the error bars relate to the uncertainty in localizing the exact position of the surface state rendered by the red boxes), compare very well with the experimental data. In particular, it reproduces the shift in the surface-state binding energy as a function of the increase of dimerization in the $(6\times1)$ phase.

Note that our calculations do not exclude the possibility that the bulk band gap hosting the surface state at $\bar{\Gamma}$ [see Fig.~\ref{Figure DFT}~(b)] closes in the low-temperature phases. Closure of the bulk band gap would mean that the surface state becomes a surface resonance, with a deeper extension of its wave function into the bulk and an enhanced sensitivity to bulk physics.

Based on these observations, we consider the energy of the surface state at the highest binding energy in our experimental ARPES data as a marker of the phase and stripe periodicities. EDCs at different temperatures are shown in Fig.~\ref{Figure UV}~(e) upon cooling (left) and warming (right) the sample. Interestingly, they exhibit a shift in energy and can be collected in different groups upon cooling, but do not show many changes upon warming. We have fitted them in the displayed energy range with a single Gaussian. The resulting surface-state energy position is displayed on Fig.~\ref{Figure UV}~(f) (full symbols) as a function of temperature. The shift in binding energy of the surface state derived from the DFT calculations in the $(6\times1)$ phase (with respect to the $(1\times1)$ case) is reported as orange dashed lines, showing a very good agreement with experimental data. \textit{Three} different sharp transitions can be observed upon cooling and, by comparison with the XPS data [Fig.~\ref{Figure XPS}~(c)], we identify the transition at $T_{c_1}$ into the $(5\times1)$ phase, at $T_{c_2}$ the transition into the $(8\times1)$ phase and at $T_{c_3}=165$~K the transition into the $(6\times1)$ phase. All of them are expected to be first-order transitions but, surprisingly, we do not observe the distinct hysteresis of the $(8\times1)$ and $(6\times1)$ phases upon warming up to $240$~K. 

\begin{figure}[h!]
\includegraphics[width=1\columnwidth]{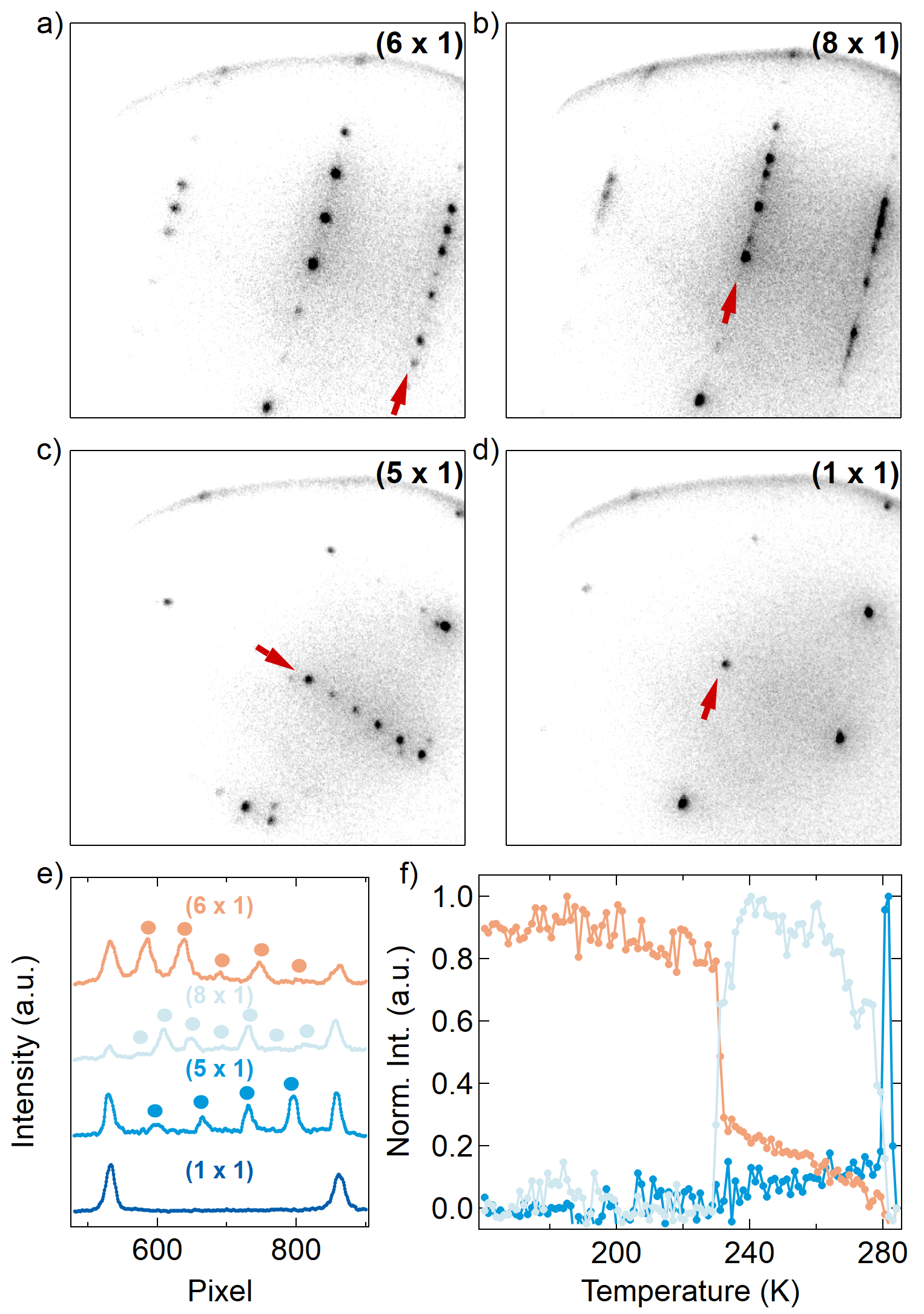}
\caption{\label{Figure LEED} 
(Color online)Raw LEED images from a region of the sample of around $20$~$\mu$m diameter in the (a) $(6\times1)$ phase at $210$~K upon warming, (b) $(8\times1)$ phase at $260$~K upon warming, (c) $(5\times1)$ phase at $280$~K upon warming, and (d) $(1\times1)$ phase at room temperature. All images obtained at 60 eV electron energy. (e) Line cuts of LEED images shown in graphs (a)$-$(d) (see the arrows indicating the position of the cuts) taken in different structural phases (at different temperatures). Solid markers highlight the superstructure peak positions in each phase. Curves are offset vertically for clarity. (f) Intensity in a single superstructure spot for the $(6\times1)$, $(8\times1)$, and $(5\times1)$ phases as a function of temperature during warming.
} 
\end{figure}

\subsection{LEED upon warming}

To investigate in more detail possible changes in the stripe phases upon warming, we have performed $\mu$LEED measurements on IrTe$_2$. 
Figures~\ref{Figure LEED}~(a)$-$\ref{Figure LEED}~(d) shows the raw LEED images taken at different temperatures during the warming process. It illustrates (a) the $(6\times1)$ phase at $210$~K, (b) the $(8\times1)$ at $260$~K, (c) the $(5\times1)$ phase at $280$~K and, finally, (d) the $(1\times1)$ phase at $295$~K. All graphs clearly show the different superstructures characterizing each phase at their respective temperatures.
Line cuts through the LEED images are shown in Fig.~\ref{Figure LEED}~(e), revealing the surface $(1\times1)$ diffraction spots and the superstructure spots corresponding to different structural phases obtained at different temperatures. In each case, only a single phase was present within the selected real-space region. IrTe$_2$ has then been measured continuously with $\mu$LEED while heating the sample, starting from the $(6\times1)$-dominated phase at about $160$~K. The evolution of the contribution of the $(6\times1)$, $(8\times1)$ and $(5\times1)$ spots to the LEED images is shown in Fig.~\ref{Figure LEED}~(f). From the lowest temperature of $160$~K, the $(6\times1)$ phase remains the only phase visible until $230$~K when the system transforms abruptly to the $(8\times1)$ phase. The $(8\times1)$ phase persists up to $280$~K and then disappears, transiently giving way to the $(5\times1)$ phase, before the $(1\times1)$ phase sets in again.

\subsection{Discussion}

We can now complete the phase diagram of Fig.~\ref{Figure UV}~(f) with the $\mu$LEED results \footnote{Formally, we do not measure electronic surface states with $\mu$LEED, but we make use of the transition temperatures (upon warming) extracted with $\mu$LEED to extrapolate the phase diagram of Fig.~\ref{Figure UV}~(f).} (red dashed line) and obtain the full picture of phase transitions occurring at the surface of IrTe$_2$, in very good agreement with previous LEED and STM works~\cite{HsuPRL,ChenPhysRev,MauererPRB,LiNatCom}. In particular, we observe that the $(6\times1)$ phase appears at $T_{c_3}=165$~K upon cooling and then persists up to $230$~K upon warming. Therefore, with our combined XPS, ARPES, and $\mu$LEED study, we clearly reveal the hysteresis of the $(6\times1)$, $(8\times1)$, and $(5\times1)$ phases over the cooling/warming cycle.

In addition to this, there is still a puzzling observation based on our XPS data [see Fig.~\ref{Figure XPS}~(c)]: Upon warming, at about $110$~K, the Ir$^{4+}$/(Ir$^{3+}$+Ir$^{4+}$) ratio jumps from a value of $0.61$ to $0.65$, indicating that the proportion of the $(6\times1)$ phase increases to nearly 100\% at this temperature, in comparison to the situation at $30$~K, for which about $30$\% of the surface is still populated by $(8\times1)$ domains. The phase change at $110$~K occurs without any shift of the related surface state energy [Fig.~\ref{Figure UV}~(f)], meaning that there is no significant concomitant structural change in the dominating $(6\times1)$ phase. Therefore we conclude that, upon warming above $110$~K, the surface of IrTe$_2$ gains sufficient thermal energy to allow the minority $(8\times1)$ domains to overcome the kinetic energy barrier to transit into $(6\times1)$ domains~\cite{MauererPRB}. This suggests that temperature cycling around $110$~K might be an efficient way to prepare a full surface $(6\times1)$ phase in IrTe$_2$. However, this complete transition to the $(6\times1)$ phase still contains incoherent domains of different orientations that result in broad ARPES spectra (see Appendix \ref{sec_App_ARPES}).

An alternative explanation is that the subsurface IrTe$_2$ layers might transit from the dominant bulk $(8\times1)$ phase (given our cooling rate, $(6\times1\times6)$ domains might also appear in the bulk \cite{TakuboXRD}) and lock-in to the surface $(6\times1)$ phase. XPS at a photon energy of $200$~eV probes electronic states deeper in the sample than ARPES at $21.22$~eV. Assuming a subsurface domain reorganization, it is interesting to note that there are hints of a weak transition occurring upon warming around $230$~K in transport data from the literature~\cite{KoNatCom,FangScienRep,Ivashko2017}, the temperature at which the $(6\times1)$ phase disappears in our data [see Fig.~\ref{Figure LEED}~(b)].

\section{Conclusion}

In this paper, we have studied the electronic structure of IrTe$_2$ with ARPES and XPS and have performed a detailed and systematic temperature dependent analysis across its charge-ordered phases. A first-order transition at $165$~K between the $(8\times1)$ and $(6\times1)$ phases is revealed. Using XPS, we have analyzed precisely the relative intensities of the Ir $4f$ core-level and related them to the dimer populations in the different charge-ordered phases. Furthermore, we have identified a particular surface-state that shifts in binding energy across the phase transitions, an observation confirmed by DFT calculations demonstrating that this surface-state binding energy is a function of the Ir dimer length. Our results therefore establish a solid basis for further photoemission studies of IrTe$_2$ under more exotic conditions like temperature quenching~\cite{OikeSuper} or time-resolved studies, or using thin samples~\cite{YoshidaSC} and, ultimately, a monolayer of IrTe$_2$.

\section{Acknowledgments} 

This project was supported from the Swiss National Science Foundation (SNSF) Grant No. P00P2\_170597. A.P. acknowledges the Osk. Huttunen Foundation for financial support, and CSC – IT Center for Science, Finland, for computational resources.  B.B. acknowledges support from the COST Action CA16218. We acknowledge the Paul Scherrer Institute, Villigen, Switzerland for provision of synchrotron radiation beamtime at beamline PEARL of the Swiss Light Source. We are very grateful to P.~Aebi for fruitful discussions and for sharing with us his photoemission setup. Skillful technical assistance was provided by F.~Bourqui, B.~Hediger and O.~Raetzo. 

\section{Appendices}

\subsection{Extended ARPES data}
\label{sec_App_ARPES}

\begin{figure}[t]
\centering
\includegraphics[width=1\columnwidth]{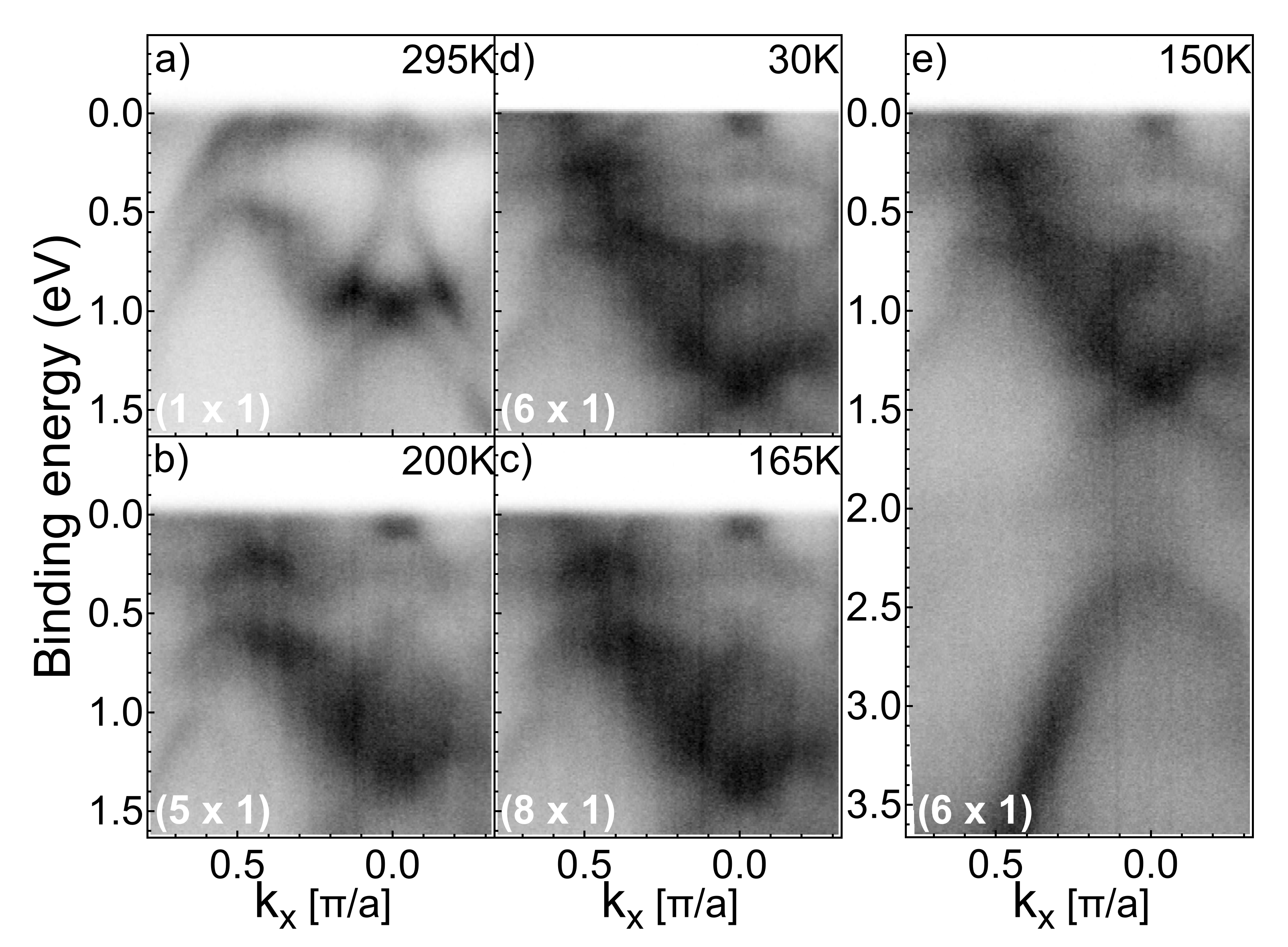}
\caption{ARPES spectra of IrTe$_2$ measured along the AL direction for $h\nu = 21.22$~eV corresponding to (a) the $(1\times1)$ phase at $295$~K , (b) the $(5\times1)$ at $200$~K, (c) the $(8\times1)$ at $165$~K, (d) the $(6\times1)$ phase at $30$~K, and (e) during the warming process at $150$~K.\label{Figure ARPES}} 
\end{figure}

Figure~\ref{Figure ARPES} shows ARPES spectra taken at different temperatures during the cooling process, (a) $295$~K, (b) $200$~K, (c) $165$~K, (d) $30$~K, and (e) at $150$~K during the warming process. The different electronic bands have already been discussed in the main text. Figure~\ref{Figure ARPES} allows us to distinguish the evolution of the electronic structure of IrTe$_2$ and, in particular, of its surface state, as a function of temperature. We can differentiate the changes of the surface state from phase $(1\times1)$, $(5\times1)$, $(8\times1)$ to phase $(6\times1)$. We can note the robustness of the $(6\times1)$ phase during the warming process, seeing only negligible changes between spectra Figs.~\ref{Figure ARPES}~(d) and \ref{Figure ARPES}~(e).

\subsection{Extended DFT data}
\label{sec_App_DFT}

The bulk energies (eV$/$atom) are $-5.307$ for the $(1\times1)$ phase, $-5.312$ for the $(5\times1)$ phase, $-5.509$ for the $(6\times1)$ phase, and $-5.511$ for the $(8\times1)$ phase. They are obtained from relaxed structures calculated with PBE functional and relaxation is done without spin-orbit interaction. In this framework, the sequence of structures from lowest to highest energy is $(8\times1)$ $\rightarrow$ $(6\times1)$ $\rightarrow$ $(5\times1)$ $\rightarrow$ $(1\times1)$. This is very close to what is observed experimentally, except for the inversion of the $(8\times1)$ and $(6\times1)$ phases. However, the $(6\times1)$ and $(8\times1)$ phases are very close in energy.

\begin{figure}[h!]
\centering
\includegraphics[width=1\columnwidth]{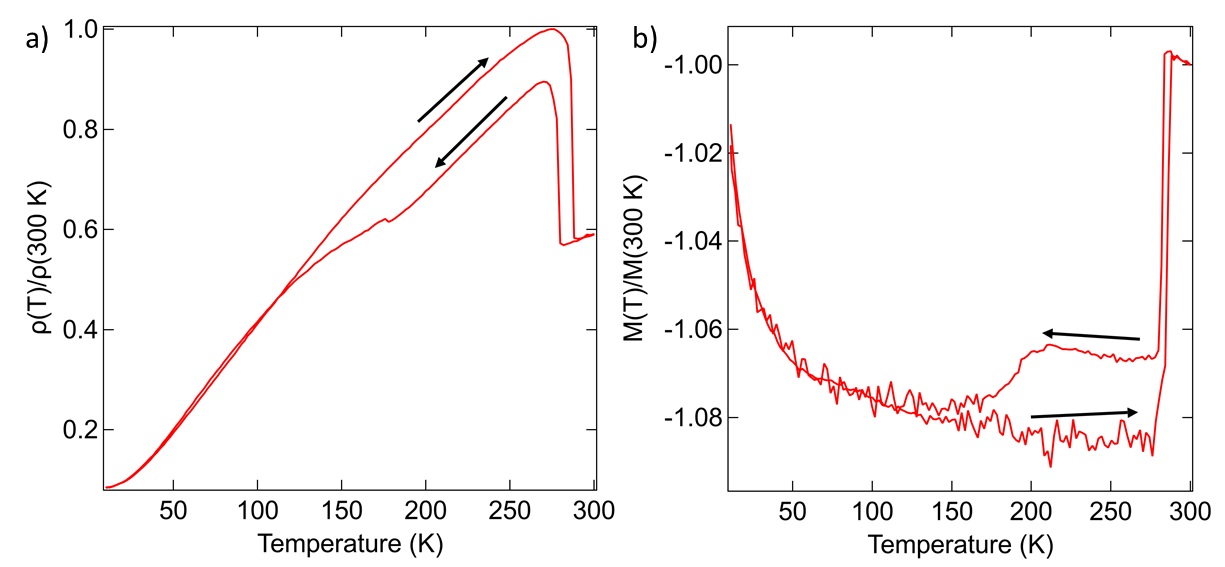}
\caption{Temperature dependence of (a) the normalized resistivity $\rho($T$)/\rho(300$~K$)$ and (b) of the magnetic susceptibility of our IrTe$_2$ single crystals.\label{Figure Transport}} 
\end{figure}

\subsection{Transport data}
\label{sec_App_transport}

In Fig.~\ref{Figure Transport}, we show the temperature-dependent [graph (a)] resistivity and [graph (b)] magnetization curves of a single crystal of IrTe$_2$ of the batch, which we used for this study. The magnetization was measured in an external field of $\mu_0H = 1$ T, with the field perpendicular to the $ab$ plane of the crystal. The electrical transport measurements were performed in a standard four-probe configuration in the $ab$ plane of the crystal. The sharp, well-defined transitions are strong indicators for the very high quality and order of the used crystals. From these measurements, we infer the critical temperatures of the charge-ordered transitions $T_{c_1}$ $= 278$~K and $T_{c_2}$ = $180$~K.


\begin{thebibliography}{49}%
\makeatletter
\providecommand \@ifxundefined [1]{%
 \@ifx{#1\undefined}
}%
\providecommand \@ifnum [1]{%
 \ifnum #1\expandafter \@firstoftwo
 \else \expandafter \@secondoftwo
 \fi
}%
\providecommand \@ifx [1]{%
 \ifx #1\expandafter \@firstoftwo
 \else \expandafter \@secondoftwo
 \fi
}%
\providecommand \natexlab [1]{#1}%
\providecommand \enquote  [1]{``#1''}%
\providecommand \bibnamefont  [1]{#1}%
\providecommand \bibfnamefont [1]{#1}%
\providecommand \citenamefont [1]{#1}%
\providecommand \href@noop [0]{\@secondoftwo}%
\providecommand \href [0]{\begingroup \@sanitize@url \@href}%
\providecommand \@href[1]{\@@startlink{#1}\@@href}%
\providecommand \@@href[1]{\endgroup#1\@@endlink}%
\providecommand \@sanitize@url [0]{\catcode `\\12\catcode `\$12\catcode
  `\&12\catcode `\#12\catcode `\^12\catcode `\_12\catcode `\%12\relax}%
\providecommand \@@startlink[1]{}%
\providecommand \@@endlink[0]{}%
\providecommand \url  [0]{\begingroup\@sanitize@url \@url }%
\providecommand \@url [1]{\endgroup\@href {#1}{\urlprefix }}%
\providecommand \urlprefix  [0]{URL }%
\providecommand \Eprint [0]{\href }%
\providecommand \doibase [0]{http://dx.doi.org/}%
\providecommand \selectlanguage [0]{\@gobble}%
\providecommand \bibinfo  [0]{\@secondoftwo}%
\providecommand \bibfield  [0]{\@secondoftwo}%
\providecommand \translation [1]{[#1]}%
\providecommand \BibitemOpen [0]{}%
\providecommand \bibitemStop [0]{}%
\providecommand \bibitemNoStop [0]{.\EOS\space}%
\providecommand \EOS [0]{\spacefactor3000\relax}%
\providecommand \BibitemShut  [1]{\csname bibitem#1\endcsname}%
\let\auto@bib@innerbib\@empty
\bibitem [{\citenamefont {Wang}\ \emph {et~al.}(2012)\citenamefont {Wang},
  \citenamefont {Kalantar-Zadeh}, \citenamefont {Kis}, \citenamefont
  {Coleman},\ and\ \citenamefont {Strano}}]{WangNat}%
  \BibitemOpen
  \bibfield  {author} {\bibinfo {author} {\bibfnamefont {Q.~H.}\ \bibnamefont
  {Wang}}, \bibinfo {author} {\bibfnamefont {K.}~\bibnamefont
  {Kalantar-Zadeh}}, \bibinfo {author} {\bibfnamefont {A.}~\bibnamefont {Kis}},
  \bibinfo {author} {\bibfnamefont {J.~N.}\ \bibnamefont {Coleman}}, \ and\
  \bibinfo {author} {\bibfnamefont {M.~S.}\ \bibnamefont {Strano}},\ }\href
  {https://doi.org/10.1038/nnano.2012.193} {\bibfield  {journal} {\bibinfo
  {journal} {Nature Nanotechnology}\ }\textbf {\bibinfo {volume} {7}},\
  \bibinfo {pages} {699} (\bibinfo {year} {2012})}\BibitemShut {NoStop}%
\bibitem [{\citenamefont {Radisavljevic}\ \emph {et~al.}(2011)\citenamefont
  {Radisavljevic}, \citenamefont {Radenovic}, \citenamefont {Brivio},
  \citenamefont {Giacometti},\ and\ \citenamefont {Kis}}]{RadisavljevicNat}%
  \BibitemOpen
  \bibfield  {author} {\bibinfo {author} {\bibfnamefont {B.}~\bibnamefont
  {Radisavljevic}}, \bibinfo {author} {\bibfnamefont {A.}~\bibnamefont
  {Radenovic}}, \bibinfo {author} {\bibfnamefont {J.}~\bibnamefont {Brivio}},
  \bibinfo {author} {\bibfnamefont {V.}~\bibnamefont {Giacometti}}, \ and\
  \bibinfo {author} {\bibfnamefont {A.}~\bibnamefont {Kis}},\ }\href {\doibase
  10.1038/nnano.2010.279} {\bibfield  {journal} {\bibinfo  {journal} {Nature
  Nanotechnology}\ }\textbf {\bibinfo {volume} {6}},\ \bibinfo {pages} {147}
  (\bibinfo {year} {2011})}\BibitemShut {NoStop}%
\bibitem [{\citenamefont {Mak}\ \emph {et~al.}(2012)\citenamefont {Mak},
  \citenamefont {He}, \citenamefont {Shan},\ and\ \citenamefont
  {Heinz}}]{MakNat}%
  \BibitemOpen
  \bibfield  {author} {\bibinfo {author} {\bibfnamefont {K.~F.}\ \bibnamefont
  {Mak}}, \bibinfo {author} {\bibfnamefont {K.}~\bibnamefont {He}}, \bibinfo
  {author} {\bibfnamefont {J.}~\bibnamefont {Shan}}, \ and\ \bibinfo {author}
  {\bibfnamefont {T.~F.}\ \bibnamefont {Heinz}},\ }\href {\doibase
  10.1038/nnano.2012.96} {\bibfield  {journal} {\bibinfo  {journal} {Nature
  Nanotechnology}\ }\textbf {\bibinfo {volume} {7}},\ \bibinfo {pages} {494}
  (\bibinfo {year} {2012})}\BibitemShut {NoStop}%
\bibitem [{\citenamefont {Bertoni}\ \emph {et~al.}(2016)\citenamefont
  {Bertoni}, \citenamefont {Nicholson}, \citenamefont {Waldecker},
  \citenamefont {H\"ubener}, \citenamefont {Monney}, \citenamefont
  {De~Giovannini}, \citenamefont {Puppin}, \citenamefont {Hoesch},
  \citenamefont {Springate}, \citenamefont {Chapman}, \citenamefont {Cacho},
  \citenamefont {Wolf}, \citenamefont {Rubio},\ and\ \citenamefont
  {Ernstorfer}}]{BertoniPRL}%
  \BibitemOpen
  \bibfield  {author} {\bibinfo {author} {\bibfnamefont {R.}~\bibnamefont
  {Bertoni}}, \bibinfo {author} {\bibfnamefont {C.~W.}\ \bibnamefont
  {Nicholson}}, \bibinfo {author} {\bibfnamefont {L.}~\bibnamefont
  {Waldecker}}, \bibinfo {author} {\bibfnamefont {H.}~\bibnamefont
  {H\"ubener}}, \bibinfo {author} {\bibfnamefont {C.}~\bibnamefont {Monney}},
  \bibinfo {author} {\bibfnamefont {U.}~\bibnamefont {De~Giovannini}}, \bibinfo
  {author} {\bibfnamefont {M.}~\bibnamefont {Puppin}}, \bibinfo {author}
  {\bibfnamefont {M.}~\bibnamefont {Hoesch}}, \bibinfo {author} {\bibfnamefont
  {E.}~\bibnamefont {Springate}}, \bibinfo {author} {\bibfnamefont {R.~T.}\
  \bibnamefont {Chapman}}, \bibinfo {author} {\bibfnamefont {C.}~\bibnamefont
  {Cacho}}, \bibinfo {author} {\bibfnamefont {M.}~\bibnamefont {Wolf}},
  \bibinfo {author} {\bibfnamefont {A.}~\bibnamefont {Rubio}}, \ and\ \bibinfo
  {author} {\bibfnamefont {R.}~\bibnamefont {Ernstorfer}},\ }\href {\doibase
  10.1103/PhysRevLett.117.277201} {\bibfield  {journal} {\bibinfo  {journal}
  {Phys. Rev. Lett.}\ }\textbf {\bibinfo {volume} {117}},\ \bibinfo {pages}
  {277201} (\bibinfo {year} {2016})}\BibitemShut {NoStop}%
\bibitem [{\citenamefont {Rossnagel}(2011)}]{RossnagelIOP}%
  \BibitemOpen
  \bibfield  {author} {\bibinfo {author} {\bibfnamefont {K.}~\bibnamefont
  {Rossnagel}},\ }\href {\doibase 10.1088/0953-8984/23/21/213001} {\bibfield
  {journal} {\bibinfo  {journal} {Journal of Physics: Condensed Matter}\
  }\textbf {\bibinfo {volume} {23}},\ \bibinfo {pages} {213001} (\bibinfo
  {year} {2011})}\BibitemShut {NoStop}%
\bibitem [{\citenamefont {Johannes}\ and\ \citenamefont
  {Mazin}(2008)}]{JohannesPRB}%
  \BibitemOpen
  \bibfield  {author} {\bibinfo {author} {\bibfnamefont {M.~D.}\ \bibnamefont
  {Johannes}}\ and\ \bibinfo {author} {\bibfnamefont {I.~I.}\ \bibnamefont
  {Mazin}},\ }\href {\doibase 10.1103/PhysRevB.77.165135} {\bibfield  {journal}
  {\bibinfo  {journal} {Phys. Rev. B}\ }\textbf {\bibinfo {volume} {77}},\
  \bibinfo {pages} {165135} (\bibinfo {year} {2008})}\BibitemShut {NoStop}%
\bibitem [{\citenamefont {Pyon}\ \emph {et~al.}(2012)\citenamefont {Pyon},
  \citenamefont {Kudo},\ and\ \citenamefont {Nohara}}]{PyonJPSJ}%
  \BibitemOpen
  \bibfield  {author} {\bibinfo {author} {\bibfnamefont {S.}~\bibnamefont
  {Pyon}}, \bibinfo {author} {\bibfnamefont {K.}~\bibnamefont {Kudo}}, \ and\
  \bibinfo {author} {\bibfnamefont {M.}~\bibnamefont {Nohara}},\ }\href
  {\doibase 10.1143/JPSJ.81.053701} {\bibfield  {journal} {\bibinfo  {journal}
  {Journal of the Physical Society of Japan}\ }\textbf {\bibinfo {volume}
  {81}},\ \bibinfo {pages} {053701} (\bibinfo {year} {2012})}\BibitemShut
  {NoStop}%
\bibitem [{\citenamefont {Chen}\ \emph {et~al.}(2016)\citenamefont {Chen},
  \citenamefont {Chan}, \citenamefont {Wong}, \citenamefont {Fang},
  \citenamefont {Chou}, \citenamefont {Mo}, \citenamefont {Hussain},
  \citenamefont {Fedorov},\ and\ \citenamefont {Chiang}}]{ChenNanoLet}%
  \BibitemOpen
  \bibfield  {author} {\bibinfo {author} {\bibfnamefont {P.}~\bibnamefont
  {Chen}}, \bibinfo {author} {\bibfnamefont {Y.-H.}\ \bibnamefont {Chan}},
  \bibinfo {author} {\bibfnamefont {M.-H.}\ \bibnamefont {Wong}}, \bibinfo
  {author} {\bibfnamefont {X.-Y.}\ \bibnamefont {Fang}}, \bibinfo {author}
  {\bibfnamefont {M.~Y.}\ \bibnamefont {Chou}}, \bibinfo {author}
  {\bibfnamefont {S.-K.}\ \bibnamefont {Mo}}, \bibinfo {author} {\bibfnamefont
  {Z.}~\bibnamefont {Hussain}}, \bibinfo {author} {\bibfnamefont {A.-V.}\
  \bibnamefont {Fedorov}}, \ and\ \bibinfo {author} {\bibfnamefont {T.-C.}\
  \bibnamefont {Chiang}},\ }\href {\doibase 10.1021/acs.nanolett.6b02710}
  {\bibfield  {journal} {\bibinfo  {journal} {Nano Letters}\ }\textbf {\bibinfo
  {volume} {16}},\ \bibinfo {pages} {6331} (\bibinfo {year}
  {2016})}\BibitemShut {NoStop}%
\bibitem [{\citenamefont {Navarro-Mortalla}\ \emph {et~al.}(2016)\citenamefont
  {Navarro-Mortalla}, \citenamefont {Island}, \citenamefont {Manas-Valero},
  \citenamefont {Pinilla-Cienfuegos}, \citenamefont {Castellanos-Gomez},
  \citenamefont {Quereda}, \citenamefont {Rubio-Bollinger}, \citenamefont
  {Chirolli}, \citenamefont {Silva-Guillen}, \citenamefont {Agraït},
  \citenamefont {Steele}, \citenamefont {Guinea}, \citenamefont {van~der
  Zant},\ and\ \citenamefont {Coronado}}]{NavarroNatCom}%
  \BibitemOpen
  \bibfield  {author} {\bibinfo {author} {\bibfnamefont {E.}~\bibnamefont
  {Navarro-Mortalla}}, \bibinfo {author} {\bibfnamefont {J.~O.}\ \bibnamefont
  {Island}}, \bibinfo {author} {\bibfnamefont {S.}~\bibnamefont
  {Manas-Valero}}, \bibinfo {author} {\bibfnamefont {E.}~\bibnamefont
  {Pinilla-Cienfuegos}}, \bibinfo {author} {\bibfnamefont {A.}~\bibnamefont
  {Castellanos-Gomez}}, \bibinfo {author} {\bibfnamefont {J.}~\bibnamefont
  {Quereda}}, \bibinfo {author} {\bibfnamefont {G.}~\bibnamefont
  {Rubio-Bollinger}}, \bibinfo {author} {\bibfnamefont {L.}~\bibnamefont
  {Chirolli}}, \bibinfo {author} {\bibfnamefont {J.~A.}\ \bibnamefont
  {Silva-Guillen}}, \bibinfo {author} {\bibfnamefont {N.}~\bibnamefont
  {Agraït}}, \bibinfo {author} {\bibfnamefont {G.~A.}\ \bibnamefont {Steele}},
  \bibinfo {author} {\bibfnamefont {F.}~\bibnamefont {Guinea}}, \bibinfo
  {author} {\bibfnamefont {H.~S.~J.}\ \bibnamefont {van~der Zant}}, \ and\
  \bibinfo {author} {\bibfnamefont {E.}~\bibnamefont {Coronado}},\ }\href
  {\doibase 10.1038/ncomms11043} {\bibfield  {journal} {\bibinfo  {journal}
  {Nature Communications}\ }\textbf {\bibinfo {volume} {7}},\ \bibinfo {pages}
  {11043} (\bibinfo {year} {2016})}\BibitemShut {NoStop}%
\bibitem [{\citenamefont {Chen}\ \emph {et~al.}(2018)\citenamefont {Chen},
  \citenamefont {Pai}, \citenamefont {Chan}, \citenamefont {Madhavan},
  \citenamefont {Chou}, \citenamefont {Mo}, \citenamefont {Fedorov},\ and\
  \citenamefont {Chiang}}]{ChenPRL}%
  \BibitemOpen
  \bibfield  {author} {\bibinfo {author} {\bibfnamefont {P.}~\bibnamefont
  {Chen}}, \bibinfo {author} {\bibfnamefont {W.~W.}\ \bibnamefont {Pai}},
  \bibinfo {author} {\bibfnamefont {Y.-H.}\ \bibnamefont {Chan}}, \bibinfo
  {author} {\bibfnamefont {V.}~\bibnamefont {Madhavan}}, \bibinfo {author}
  {\bibfnamefont {M.~Y.}\ \bibnamefont {Chou}}, \bibinfo {author}
  {\bibfnamefont {S.-K.}\ \bibnamefont {Mo}}, \bibinfo {author} {\bibfnamefont
  {A.-V.}\ \bibnamefont {Fedorov}}, \ and\ \bibinfo {author} {\bibfnamefont
  {T.-C.}\ \bibnamefont {Chiang}},\ }\href {\doibase
  10.1103/PhysRevLett.121.196402} {\bibfield  {journal} {\bibinfo  {journal}
  {Phys. Rev. Lett.}\ }\textbf {\bibinfo {volume} {121}},\ \bibinfo {pages}
  {196402} (\bibinfo {year} {2018})}\BibitemShut {NoStop}%
\bibitem [{\citenamefont {Pascut}\ \emph
  {et~al.}(2014{\natexlab{a}})\citenamefont {Pascut}, \citenamefont {Haule},
  \citenamefont {Gutmann}, \citenamefont {Barnett}, \citenamefont {Bombardi},
  \citenamefont {Artyukhin}, \citenamefont {Birol}, \citenamefont {Vanderbilt},
  \citenamefont {Yang}, \citenamefont {Cheong},\ and\ \citenamefont
  {Kiryukhin}}]{PascutPRL}%
  \BibitemOpen
  \bibfield  {author} {\bibinfo {author} {\bibfnamefont {G.~L.}\ \bibnamefont
  {Pascut}}, \bibinfo {author} {\bibfnamefont {K.}~\bibnamefont {Haule}},
  \bibinfo {author} {\bibfnamefont {M.~J.}\ \bibnamefont {Gutmann}}, \bibinfo
  {author} {\bibfnamefont {S.~A.}\ \bibnamefont {Barnett}}, \bibinfo {author}
  {\bibfnamefont {A.}~\bibnamefont {Bombardi}}, \bibinfo {author}
  {\bibfnamefont {S.}~\bibnamefont {Artyukhin}}, \bibinfo {author}
  {\bibfnamefont {T.}~\bibnamefont {Birol}}, \bibinfo {author} {\bibfnamefont
  {D.}~\bibnamefont {Vanderbilt}}, \bibinfo {author} {\bibfnamefont {J.~J.}\
  \bibnamefont {Yang}}, \bibinfo {author} {\bibfnamefont {S.-W.}\ \bibnamefont
  {Cheong}}, \ and\ \bibinfo {author} {\bibfnamefont {V.}~\bibnamefont
  {Kiryukhin}},\ }\href {\doibase 10.1103/PhysRevLett.112.086402} {\bibfield
  {journal} {\bibinfo  {journal} {Phys. Rev. Lett.}\ }\textbf {\bibinfo
  {volume} {112}},\ \bibinfo {pages} {086402} (\bibinfo {year}
  {2014}{\natexlab{a}})}\BibitemShut {NoStop}%
\bibitem [{\citenamefont {Mauerer}\ \emph {et~al.}(2016)\citenamefont
  {Mauerer}, \citenamefont {Vogt}, \citenamefont {Hsu}, \citenamefont {Pascut},
  \citenamefont {Haule}, \citenamefont {Kiryukhin}, \citenamefont {Yang},
  \citenamefont {Cheong}, \citenamefont {Wu},\ and\ \citenamefont
  {Bode}}]{MauererPRB}%
  \BibitemOpen
  \bibfield  {author} {\bibinfo {author} {\bibfnamefont {T.}~\bibnamefont
  {Mauerer}}, \bibinfo {author} {\bibfnamefont {M.}~\bibnamefont {Vogt}},
  \bibinfo {author} {\bibfnamefont {P.-J.}\ \bibnamefont {Hsu}}, \bibinfo
  {author} {\bibfnamefont {G.~L.}\ \bibnamefont {Pascut}}, \bibinfo {author}
  {\bibfnamefont {K.}~\bibnamefont {Haule}}, \bibinfo {author} {\bibfnamefont
  {V.}~\bibnamefont {Kiryukhin}}, \bibinfo {author} {\bibfnamefont
  {J.}~\bibnamefont {Yang}}, \bibinfo {author} {\bibfnamefont {S.-W.}\
  \bibnamefont {Cheong}}, \bibinfo {author} {\bibfnamefont {W.}~\bibnamefont
  {Wu}}, \ and\ \bibinfo {author} {\bibfnamefont {M.}~\bibnamefont {Bode}},\
  }\href {\doibase 10.1103/PhysRevB.94.014106} {\bibfield  {journal} {\bibinfo
  {journal} {Phys. Rev. B}\ }\textbf {\bibinfo {volume} {94}},\ \bibinfo
  {pages} {014106} (\bibinfo {year} {2016})}\BibitemShut {NoStop}%
\bibitem [{\citenamefont {Hsu}\ \emph {et~al.}(2013)\citenamefont {Hsu},
  \citenamefont {Mauerer}, \citenamefont {Vogt}, \citenamefont {Yang},
  \citenamefont {Oh}, \citenamefont {Cheong}, \citenamefont {Bode},\ and\
  \citenamefont {Wu}}]{HsuPRL}%
  \BibitemOpen
  \bibfield  {author} {\bibinfo {author} {\bibfnamefont {P.-J.}\ \bibnamefont
  {Hsu}}, \bibinfo {author} {\bibfnamefont {T.}~\bibnamefont {Mauerer}},
  \bibinfo {author} {\bibfnamefont {M.}~\bibnamefont {Vogt}}, \bibinfo {author}
  {\bibfnamefont {J.~J.}\ \bibnamefont {Yang}}, \bibinfo {author}
  {\bibfnamefont {Y.~S.}\ \bibnamefont {Oh}}, \bibinfo {author} {\bibfnamefont
  {S.-W.}\ \bibnamefont {Cheong}}, \bibinfo {author} {\bibfnamefont
  {M.}~\bibnamefont {Bode}}, \ and\ \bibinfo {author} {\bibfnamefont
  {W.}~\bibnamefont {Wu}},\ }\href {\doibase 10.1103/PhysRevLett.111.266401}
  {\bibfield  {journal} {\bibinfo  {journal} {Phys. Rev. Lett.}\ }\textbf
  {\bibinfo {volume} {111}},\ \bibinfo {pages} {266401} (\bibinfo {year}
  {2013})}\BibitemShut {NoStop}%
\bibitem [{\citenamefont {Yoshida}\ \emph {et~al.}(2018)\citenamefont
  {Yoshida}, \citenamefont {Kudo}, \citenamefont {Nohara},\ and\ \citenamefont
  {Iwasa}}]{YoshidaSC}%
  \BibitemOpen
  \bibfield  {author} {\bibinfo {author} {\bibfnamefont {M.}~\bibnamefont
  {Yoshida}}, \bibinfo {author} {\bibfnamefont {K.}~\bibnamefont {Kudo}},
  \bibinfo {author} {\bibfnamefont {M.}~\bibnamefont {Nohara}}, \ and\ \bibinfo
  {author} {\bibfnamefont {Y.}~\bibnamefont {Iwasa}},\ }\href
  {https://doi.org/10.1021/acs.nanolett.8b00673} {\bibfield  {journal}
  {\bibinfo  {journal} {Nano Letters}\ }\textbf {\bibinfo {volume} {18}},\
  \bibinfo {pages} {3113} (\bibinfo {year} {2018})}\BibitemShut {NoStop}%
\bibitem [{\citenamefont {Oike}\ \emph {et~al.}(2018)\citenamefont {Oike},
  \citenamefont {Kamitani}, \citenamefont {Tokura},\ and\ \citenamefont
  {Kagawa}}]{OikeSuper}%
  \BibitemOpen
  \bibfield  {author} {\bibinfo {author} {\bibfnamefont {H.}~\bibnamefont
  {Oike}}, \bibinfo {author} {\bibfnamefont {M.}~\bibnamefont {Kamitani}},
  \bibinfo {author} {\bibfnamefont {Y.}~\bibnamefont {Tokura}}, \ and\ \bibinfo
  {author} {\bibfnamefont {F.}~\bibnamefont {Kagawa}},\ }\href {\doibase
  10.1126/sciadv.aau3489} {\bibfield  {journal} {\bibinfo  {journal} {Science
  Advances}\ }\textbf {\bibinfo {volume} {4}} (\bibinfo {year} {2018}),\
  10.1126/sciadv.aau3489}\BibitemShut {NoStop}%
\bibitem [{\citenamefont {Ko}\ \emph {et~al.}(2015)\citenamefont {Ko},
  \citenamefont {Lee}, \citenamefont {Kim}, \citenamefont {Yang}, \citenamefont
  {Cheong}, \citenamefont {Eom}, \citenamefont {Kim}, \citenamefont {Gammag},
  \citenamefont {Kim}, \citenamefont {Kim}, \citenamefont {Kim}, \citenamefont
  {Yeom}, \citenamefont {Koo}, \citenamefont {Kim},\ and\ \citenamefont
  {Park}}]{KoNatCom}%
  \BibitemOpen
  \bibfield  {author} {\bibinfo {author} {\bibfnamefont {K.-T.}\ \bibnamefont
  {Ko}}, \bibinfo {author} {\bibfnamefont {H.-H.}\ \bibnamefont {Lee}},
  \bibinfo {author} {\bibfnamefont {D.-H.}\ \bibnamefont {Kim}}, \bibinfo
  {author} {\bibfnamefont {J.-J.}\ \bibnamefont {Yang}}, \bibinfo {author}
  {\bibfnamefont {S.-W.}\ \bibnamefont {Cheong}}, \bibinfo {author}
  {\bibfnamefont {M.~J.}\ \bibnamefont {Eom}}, \bibinfo {author} {\bibfnamefont
  {J.~S.}\ \bibnamefont {Kim}}, \bibinfo {author} {\bibfnamefont
  {R.}~\bibnamefont {Gammag}}, \bibinfo {author} {\bibfnamefont {K.-S.}\
  \bibnamefont {Kim}}, \bibinfo {author} {\bibfnamefont {H.-S.}\ \bibnamefont
  {Kim}}, \bibinfo {author} {\bibfnamefont {T.-H.}\ \bibnamefont {Kim}},
  \bibinfo {author} {\bibfnamefont {H.-W.}\ \bibnamefont {Yeom}}, \bibinfo
  {author} {\bibfnamefont {T.-Y.}\ \bibnamefont {Koo}}, \bibinfo {author}
  {\bibfnamefont {H.-D.}\ \bibnamefont {Kim}}, \ and\ \bibinfo {author}
  {\bibfnamefont {J.-H.}\ \bibnamefont {Park}},\ }\href {\doibase
  10.1038/ncomms8342} {\bibfield  {journal} {\bibinfo  {journal} {Nature
  Communications}\ }\textbf {\bibinfo {volume} {6}},\ \bibinfo {pages} {7342}
  (\bibinfo {year} {2015})}\BibitemShut {NoStop}%
\bibitem [{\citenamefont {Fang}\ \emph {et~al.}(2013)\citenamefont {Fang},
  \citenamefont {Xu}, \citenamefont {Dong}, \citenamefont {Zheng},\ and\
  \citenamefont {Wang}}]{FangScienRep}%
  \BibitemOpen
  \bibfield  {author} {\bibinfo {author} {\bibfnamefont {A.~F.}\ \bibnamefont
  {Fang}}, \bibinfo {author} {\bibfnamefont {G.}~\bibnamefont {Xu}}, \bibinfo
  {author} {\bibfnamefont {T.}~\bibnamefont {Dong}}, \bibinfo {author}
  {\bibfnamefont {P.}~\bibnamefont {Zheng}}, \ and\ \bibinfo {author}
  {\bibfnamefont {N.~L.}\ \bibnamefont {Wang}},\ }\href
  {https://doi.org/10.1038/srep01153} {\bibfield  {journal} {\bibinfo
  {journal} {Scientific Reports}\ }\textbf {\bibinfo {volume} {3}},\ \bibinfo
  {pages} {1153} (\bibinfo {year} {2013})}\BibitemShut {NoStop}%
\bibitem [{\citenamefont {Jobic}\ \emph {et~al.}(1991)\citenamefont {Jobic},
  \citenamefont {Deniard}, \citenamefont {Brec}, \citenamefont {Rouxel},
  \citenamefont {Jouanneaux},\ and\ \citenamefont {Fitch}}]{JobicZeit}%
  \BibitemOpen
  \bibfield  {author} {\bibinfo {author} {\bibfnamefont {S.}~\bibnamefont
  {Jobic}}, \bibinfo {author} {\bibfnamefont {P.}~\bibnamefont {Deniard}},
  \bibinfo {author} {\bibfnamefont {R.}~\bibnamefont {Brec}}, \bibinfo {author}
  {\bibfnamefont {J.}~\bibnamefont {Rouxel}}, \bibinfo {author} {\bibfnamefont
  {A.}~\bibnamefont {Jouanneaux}}, \ and\ \bibinfo {author} {\bibfnamefont
  {A.~N.}\ \bibnamefont {Fitch}},\ }\href {\doibase 10.1002/zaac.19915980119}
  {\bibfield  {journal} {\bibinfo  {journal} {Zeitschrift für anorganische und
  allgemeine Chemie}\ }\textbf {\bibinfo {volume} {598}},\ \bibinfo {pages}
  {199} (\bibinfo {year} {1991})}\BibitemShut {NoStop}%
\bibitem [{\citenamefont {Matsumoto}\ \emph {et~al.}(1999)\citenamefont
  {Matsumoto}, \citenamefont {Taniguchi}, \citenamefont {Endoh}, \citenamefont
  {Takano},\ and\ \citenamefont {Nagata}}]{MatsumotoJLTP}%
  \BibitemOpen
  \bibfield  {author} {\bibinfo {author} {\bibfnamefont {N.}~\bibnamefont
  {Matsumoto}}, \bibinfo {author} {\bibfnamefont {K.}~\bibnamefont
  {Taniguchi}}, \bibinfo {author} {\bibfnamefont {R.}~\bibnamefont {Endoh}},
  \bibinfo {author} {\bibfnamefont {H.}~\bibnamefont {Takano}}, \ and\ \bibinfo
  {author} {\bibfnamefont {S.}~\bibnamefont {Nagata}},\ }\href {\doibase
  10.1023/A:1022546928480} {\bibfield  {journal} {\bibinfo  {journal} {Journal
  of Low Temperature Physics}\ }\textbf {\bibinfo {volume} {117}},\ \bibinfo
  {pages} {1129} (\bibinfo {year} {1999})}\BibitemShut {NoStop}%
\bibitem [{\citenamefont {Toriyama}\ \emph {et~al.}(2014)\citenamefont
  {Toriyama}, \citenamefont {Kobori}, \citenamefont {Konishi}, \citenamefont
  {Ohta}, \citenamefont {Sugimoto}, \citenamefont {Kim}, \citenamefont
  {Fujiwara}, \citenamefont {Pyon}, \citenamefont {Kudo},\ and\ \citenamefont
  {Nohara}}]{ToriyamaJapanLetters}%
  \BibitemOpen
  \bibfield  {author} {\bibinfo {author} {\bibfnamefont {T.}~\bibnamefont
  {Toriyama}}, \bibinfo {author} {\bibfnamefont {M.}~\bibnamefont {Kobori}},
  \bibinfo {author} {\bibfnamefont {T.}~\bibnamefont {Konishi}}, \bibinfo
  {author} {\bibfnamefont {Y.}~\bibnamefont {Ohta}}, \bibinfo {author}
  {\bibfnamefont {K.}~\bibnamefont {Sugimoto}}, \bibinfo {author}
  {\bibfnamefont {J.}~\bibnamefont {Kim}}, \bibinfo {author} {\bibfnamefont
  {A.}~\bibnamefont {Fujiwara}}, \bibinfo {author} {\bibfnamefont
  {S.}~\bibnamefont {Pyon}}, \bibinfo {author} {\bibfnamefont {K.}~\bibnamefont
  {Kudo}}, \ and\ \bibinfo {author} {\bibfnamefont {M.}~\bibnamefont
  {Nohara}},\ }\href {\doibase 10.7566/JPSJ.83.033701} {\bibfield  {journal}
  {\bibinfo  {journal} {Journal of the Physical Society of Japan}\ }\textbf
  {\bibinfo {volume} {83}},\ \bibinfo {pages} {033701} (\bibinfo {year}
  {2014})}\BibitemShut {NoStop}%
\bibitem [{\citenamefont {Koley}(2016)}]{KoleySSCom}%
  \BibitemOpen
  \bibfield  {author} {\bibinfo {author} {\bibfnamefont {S.}~\bibnamefont
  {Koley}},\ }\href {\doibase https://doi.org/10.1016/j.ssc.2016.08.015}
  {\bibfield  {journal} {\bibinfo  {journal} {Solid State Communications}\
  }\textbf {\bibinfo {volume} {247}},\ \bibinfo {pages} {40 } (\bibinfo {year}
  {2016})}\BibitemShut {NoStop}%
\bibitem [{\citenamefont {Li}\ \emph {et~al.}(2014{\natexlab{a}})\citenamefont
  {Li}, \citenamefont {Huang}, \citenamefont {Sun},\ and\ \citenamefont
  {Xing}}]{LiSciRep}%
  \BibitemOpen
  \bibfield  {author} {\bibinfo {author} {\bibfnamefont {B.}~\bibnamefont
  {Li}}, \bibinfo {author} {\bibfnamefont {G.}~\bibnamefont {Huang}}, \bibinfo
  {author} {\bibfnamefont {J.}~\bibnamefont {Sun}}, \ and\ \bibinfo {author}
  {\bibfnamefont {Z.}~\bibnamefont {Xing}},\ }\href
  {https://doi.org/10.1038/srep06433} {\bibfield  {journal} {\bibinfo
  {journal} {Scientific Reports}\ }\textbf {\bibinfo {volume} {4}},\ \bibinfo
  {pages} {6433} (\bibinfo {year} {2014}{\natexlab{a}})}\BibitemShut {NoStop}%
\bibitem [{\citenamefont {Paris}\ \emph {et~al.}(2016)\citenamefont {Paris},
  \citenamefont {Joseph}, \citenamefont {Iadecola}, \citenamefont {Marini},
  \citenamefont {Ishii}, \citenamefont {Kudo}, \citenamefont {Pascarelli},
  \citenamefont {Nohara}, \citenamefont {Mizokawa},\ and\ \citenamefont
  {Saini}}]{ParisXAS}%
  \BibitemOpen
  \bibfield  {author} {\bibinfo {author} {\bibfnamefont {E.}~\bibnamefont
  {Paris}}, \bibinfo {author} {\bibfnamefont {B.}~\bibnamefont {Joseph}},
  \bibinfo {author} {\bibfnamefont {A.}~\bibnamefont {Iadecola}}, \bibinfo
  {author} {\bibfnamefont {C.}~\bibnamefont {Marini}}, \bibinfo {author}
  {\bibfnamefont {H.}~\bibnamefont {Ishii}}, \bibinfo {author} {\bibfnamefont
  {K.}~\bibnamefont {Kudo}}, \bibinfo {author} {\bibfnamefont {S.}~\bibnamefont
  {Pascarelli}}, \bibinfo {author} {\bibfnamefont {M.}~\bibnamefont {Nohara}},
  \bibinfo {author} {\bibfnamefont {T.}~\bibnamefont {Mizokawa}}, \ and\
  \bibinfo {author} {\bibfnamefont {N.~L.}\ \bibnamefont {Saini}},\ }\href
  {\doibase 10.1103/PhysRevB.93.134109} {\bibfield  {journal} {\bibinfo
  {journal} {Phys. Rev. B}\ }\textbf {\bibinfo {volume} {93}},\ \bibinfo
  {pages} {134109} (\bibinfo {year} {2016})}\BibitemShut {NoStop}%
\bibitem [{\citenamefont {Pascut}\ \emph
  {et~al.}(2014{\natexlab{b}})\citenamefont {Pascut}, \citenamefont {Birol},
  \citenamefont {Gutmann}, \citenamefont {Yang}, \citenamefont {Cheong},
  \citenamefont {Haule},\ and\ \citenamefont {Kiryukhin}}]{PascutPRB}%
  \BibitemOpen
  \bibfield  {author} {\bibinfo {author} {\bibfnamefont {G.~L.}\ \bibnamefont
  {Pascut}}, \bibinfo {author} {\bibfnamefont {T.}~\bibnamefont {Birol}},
  \bibinfo {author} {\bibfnamefont {M.~J.}\ \bibnamefont {Gutmann}}, \bibinfo
  {author} {\bibfnamefont {J.~J.}\ \bibnamefont {Yang}}, \bibinfo {author}
  {\bibfnamefont {S.-W.}\ \bibnamefont {Cheong}}, \bibinfo {author}
  {\bibfnamefont {K.}~\bibnamefont {Haule}}, \ and\ \bibinfo {author}
  {\bibfnamefont {V.}~\bibnamefont {Kiryukhin}},\ }\href {\doibase
  10.1103/PhysRevB.90.195122} {\bibfield  {journal} {\bibinfo  {journal} {Phys.
  Rev. B}\ }\textbf {\bibinfo {volume} {90}},\ \bibinfo {pages} {195122}
  (\bibinfo {year} {2014}{\natexlab{b}})}\BibitemShut {NoStop}%
\bibitem [{\citenamefont {Li}\ \emph {et~al.}(2014{\natexlab{b}})\citenamefont
  {Li}, \citenamefont {Lin}, \citenamefont {Yan}, \citenamefont {Chen},
  \citenamefont {Gianfrancesco}, \citenamefont {Singh}, \citenamefont
  {Mandrus}, \citenamefont {Kalinin},\ and\ \citenamefont {Pan}}]{LiNatCom}%
  \BibitemOpen
  \bibfield  {author} {\bibinfo {author} {\bibfnamefont {Q.}~\bibnamefont
  {Li}}, \bibinfo {author} {\bibfnamefont {W.}~\bibnamefont {Lin}}, \bibinfo
  {author} {\bibfnamefont {J.}~\bibnamefont {Yan}}, \bibinfo {author}
  {\bibfnamefont {X.}~\bibnamefont {Chen}}, \bibinfo {author} {\bibfnamefont
  {A.~G.}\ \bibnamefont {Gianfrancesco}}, \bibinfo {author} {\bibfnamefont
  {D.~J.}\ \bibnamefont {Singh}}, \bibinfo {author} {\bibfnamefont
  {D.}~\bibnamefont {Mandrus}}, \bibinfo {author} {\bibfnamefont {S.~V.}\
  \bibnamefont {Kalinin}}, \ and\ \bibinfo {author} {\bibfnamefont
  {M.}~\bibnamefont {Pan}},\ }\href {\doibase 10.1038/ncomms6358} {\bibfield
  {journal} {\bibinfo  {journal} {Nature Communications}\ }\textbf {\bibinfo
  {volume} {5}},\ \bibinfo {pages} {5358} (\bibinfo {year}
  {2014}{\natexlab{b}})}\BibitemShut {NoStop}%
\bibitem [{\citenamefont {Oh}\ \emph {et~al.}(2013)\citenamefont {Oh},
  \citenamefont {Yang}, \citenamefont {Horibe},\ and\ \citenamefont
  {Cheong}}]{OhPRL}%
  \BibitemOpen
  \bibfield  {author} {\bibinfo {author} {\bibfnamefont {Y.~S.}\ \bibnamefont
  {Oh}}, \bibinfo {author} {\bibfnamefont {J.~J.}\ \bibnamefont {Yang}},
  \bibinfo {author} {\bibfnamefont {Y.}~\bibnamefont {Horibe}}, \ and\ \bibinfo
  {author} {\bibfnamefont {S.-W.}\ \bibnamefont {Cheong}},\ }\href {\doibase
  10.1103/PhysRevLett.110.127209} {\bibfield  {journal} {\bibinfo  {journal}
  {Phys. Rev. Lett.}\ }\textbf {\bibinfo {volume} {110}},\ \bibinfo {pages}
  {127209} (\bibinfo {year} {2013})}\BibitemShut {NoStop}%
\bibitem [{\citenamefont {Takubo}\ \emph {et~al.}(2014)\citenamefont {Takubo},
  \citenamefont {Comin}, \citenamefont {Ootsuki}, \citenamefont {Mizokawa},
  \citenamefont {Wadati}, \citenamefont {Takahashi}, \citenamefont {Shibata},
  \citenamefont {Fujimori}, \citenamefont {Sutarto}, \citenamefont {He},
  \citenamefont {Pyon}, \citenamefont {Kudo}, \citenamefont {Nohara},
  \citenamefont {Levy}, \citenamefont {Elfimov}, \citenamefont {Sawatzky},\
  and\ \citenamefont {Damascelli}}]{TakuboPRB}%
  \BibitemOpen
  \bibfield  {author} {\bibinfo {author} {\bibfnamefont {K.}~\bibnamefont
  {Takubo}}, \bibinfo {author} {\bibfnamefont {R.}~\bibnamefont {Comin}},
  \bibinfo {author} {\bibfnamefont {D.}~\bibnamefont {Ootsuki}}, \bibinfo
  {author} {\bibfnamefont {T.}~\bibnamefont {Mizokawa}}, \bibinfo {author}
  {\bibfnamefont {H.}~\bibnamefont {Wadati}}, \bibinfo {author} {\bibfnamefont
  {Y.}~\bibnamefont {Takahashi}}, \bibinfo {author} {\bibfnamefont
  {G.}~\bibnamefont {Shibata}}, \bibinfo {author} {\bibfnamefont
  {A.}~\bibnamefont {Fujimori}}, \bibinfo {author} {\bibfnamefont
  {R.}~\bibnamefont {Sutarto}}, \bibinfo {author} {\bibfnamefont
  {F.}~\bibnamefont {He}}, \bibinfo {author} {\bibfnamefont {S.}~\bibnamefont
  {Pyon}}, \bibinfo {author} {\bibfnamefont {K.}~\bibnamefont {Kudo}}, \bibinfo
  {author} {\bibfnamefont {M.}~\bibnamefont {Nohara}}, \bibinfo {author}
  {\bibfnamefont {G.}~\bibnamefont {Levy}}, \bibinfo {author} {\bibfnamefont
  {I.~S.}\ \bibnamefont {Elfimov}}, \bibinfo {author} {\bibfnamefont {G.~A.}\
  \bibnamefont {Sawatzky}}, \ and\ \bibinfo {author} {\bibfnamefont
  {A.}~\bibnamefont {Damascelli}},\ }\href {\doibase
  10.1103/PhysRevB.90.081104} {\bibfield  {journal} {\bibinfo  {journal} {Phys.
  Rev. B}\ }\textbf {\bibinfo {volume} {90}},\ \bibinfo {pages} {081104}
  (\bibinfo {year} {2014})}\BibitemShut {NoStop}%
\bibitem [{\citenamefont {Dai}\ \emph {et~al.}(2014)\citenamefont {Dai},
  \citenamefont {Haule}, \citenamefont {Yang}, \citenamefont {Oh},
  \citenamefont {Cheong},\ and\ \citenamefont {Wu}}]{DaiPRB}%
  \BibitemOpen
  \bibfield  {author} {\bibinfo {author} {\bibfnamefont {J.}~\bibnamefont
  {Dai}}, \bibinfo {author} {\bibfnamefont {K.}~\bibnamefont {Haule}}, \bibinfo
  {author} {\bibfnamefont {J.~J.}\ \bibnamefont {Yang}}, \bibinfo {author}
  {\bibfnamefont {Y.~S.}\ \bibnamefont {Oh}}, \bibinfo {author} {\bibfnamefont
  {S.-W.}\ \bibnamefont {Cheong}}, \ and\ \bibinfo {author} {\bibfnamefont
  {W.}~\bibnamefont {Wu}},\ }\href {\doibase 10.1103/PhysRevB.90.235121}
  {\bibfield  {journal} {\bibinfo  {journal} {Phys. Rev. B}\ }\textbf {\bibinfo
  {volume} {90}},\ \bibinfo {pages} {235121} (\bibinfo {year}
  {2014})}\BibitemShut {NoStop}%
\bibitem [{\citenamefont {Chen}\ \emph {et~al.}(2017)\citenamefont {Chen},
  \citenamefont {Kim}, \citenamefont {Yang}, \citenamefont {Cao}, \citenamefont
  {Jin},\ and\ \citenamefont {Plummer}}]{ChenPhysRev}%
  \BibitemOpen
  \bibfield  {author} {\bibinfo {author} {\bibfnamefont {C.}~\bibnamefont
  {Chen}}, \bibinfo {author} {\bibfnamefont {J.}~\bibnamefont {Kim}}, \bibinfo
  {author} {\bibfnamefont {Y.}~\bibnamefont {Yang}}, \bibinfo {author}
  {\bibfnamefont {G.}~\bibnamefont {Cao}}, \bibinfo {author} {\bibfnamefont
  {R.}~\bibnamefont {Jin}}, \ and\ \bibinfo {author} {\bibfnamefont {E.~W.}\
  \bibnamefont {Plummer}},\ }\href {\doibase 10.1103/PhysRevB.95.094118}
  {\bibfield  {journal} {\bibinfo  {journal} {Phys. Rev. B}\ }\textbf {\bibinfo
  {volume} {95}},\ \bibinfo {pages} {094118} (\bibinfo {year}
  {2017})}\BibitemShut {NoStop}%
\bibitem [{\citenamefont {Takubo}\ \emph {et~al.}(2018)\citenamefont {Takubo},
  \citenamefont {Yamamoto}, \citenamefont {Hirata}, \citenamefont {Wadati},
  \citenamefont {Mizokawa}, \citenamefont {Sutarto}, \citenamefont {He},
  \citenamefont {Ishii}, \citenamefont {Yamasaki}, \citenamefont {Nakao},
  \citenamefont {Murakami}, \citenamefont {Matsuo}, \citenamefont {Ishii},
  \citenamefont {Kobayashi}, \citenamefont {Kudo},\ and\ \citenamefont
  {Nohara}}]{TakuboXRD}%
  \BibitemOpen
  \bibfield  {author} {\bibinfo {author} {\bibfnamefont {K.}~\bibnamefont
  {Takubo}}, \bibinfo {author} {\bibfnamefont {K.}~\bibnamefont {Yamamoto}},
  \bibinfo {author} {\bibfnamefont {Y.}~\bibnamefont {Hirata}}, \bibinfo
  {author} {\bibfnamefont {H.}~\bibnamefont {Wadati}}, \bibinfo {author}
  {\bibfnamefont {T.}~\bibnamefont {Mizokawa}}, \bibinfo {author}
  {\bibfnamefont {R.}~\bibnamefont {Sutarto}}, \bibinfo {author} {\bibfnamefont
  {F.}~\bibnamefont {He}}, \bibinfo {author} {\bibfnamefont {K.}~\bibnamefont
  {Ishii}}, \bibinfo {author} {\bibfnamefont {Y.}~\bibnamefont {Yamasaki}},
  \bibinfo {author} {\bibfnamefont {H.}~\bibnamefont {Nakao}}, \bibinfo
  {author} {\bibfnamefont {Y.}~\bibnamefont {Murakami}}, \bibinfo {author}
  {\bibfnamefont {G.}~\bibnamefont {Matsuo}}, \bibinfo {author} {\bibfnamefont
  {H.}~\bibnamefont {Ishii}}, \bibinfo {author} {\bibfnamefont
  {M.}~\bibnamefont {Kobayashi}}, \bibinfo {author} {\bibfnamefont
  {K.}~\bibnamefont {Kudo}}, \ and\ \bibinfo {author} {\bibfnamefont
  {M.}~\bibnamefont {Nohara}},\ }\href {\doibase 10.1103/PhysRevB.97.205142}
  {\bibfield  {journal} {\bibinfo  {journal} {Phys. Rev. B}\ }\textbf {\bibinfo
  {volume} {97}},\ \bibinfo {pages} {205142} (\bibinfo {year}
  {2018})}\BibitemShut {NoStop}%
\bibitem [{\citenamefont {Monney}\ \emph {et~al.}(2018)\citenamefont {Monney},
  \citenamefont {Schuler}, \citenamefont {Jaouen}, \citenamefont {Mottas},
  \citenamefont {Wolf}, \citenamefont {Merz}, \citenamefont {Muntwiler},
  \citenamefont {Castiglioni}, \citenamefont {Aebi}, \citenamefont {Weber},\
  and\ \citenamefont {Hengsberger}}]{MonneyPRB}%
  \BibitemOpen
  \bibfield  {author} {\bibinfo {author} {\bibfnamefont {C.}~\bibnamefont
  {Monney}}, \bibinfo {author} {\bibfnamefont {A.}~\bibnamefont {Schuler}},
  \bibinfo {author} {\bibfnamefont {T.}~\bibnamefont {Jaouen}}, \bibinfo
  {author} {\bibfnamefont {M.-L.}\ \bibnamefont {Mottas}}, \bibinfo {author}
  {\bibfnamefont {T.}~\bibnamefont {Wolf}}, \bibinfo {author} {\bibfnamefont
  {M.}~\bibnamefont {Merz}}, \bibinfo {author} {\bibfnamefont {M.}~\bibnamefont
  {Muntwiler}}, \bibinfo {author} {\bibfnamefont {L.}~\bibnamefont
  {Castiglioni}}, \bibinfo {author} {\bibfnamefont {P.}~\bibnamefont {Aebi}},
  \bibinfo {author} {\bibfnamefont {F.}~\bibnamefont {Weber}}, \ and\ \bibinfo
  {author} {\bibfnamefont {M.}~\bibnamefont {Hengsberger}},\ }\href {\doibase
  10.1103/PhysRevB.97.075110} {\bibfield  {journal} {\bibinfo  {journal} {Phys.
  Rev. B}\ }\textbf {\bibinfo {volume} {97}},\ \bibinfo {pages} {075110}
  (\bibinfo {year} {2018})}\BibitemShut {NoStop}%
\bibitem [{\citenamefont {Qian}\ \emph {et~al.}(2014)\citenamefont {Qian},
  \citenamefont {Miao}, \citenamefont {Wang}, \citenamefont {Shi},
  \citenamefont {Huang}, \citenamefont {Zhang}, \citenamefont {Xu},
  \citenamefont {Zeng}, \citenamefont {Ma}, \citenamefont {Richard},
  \citenamefont {Shi}, \citenamefont {Xu}, \citenamefont {Dai}, \citenamefont
  {Fang}, \citenamefont {Fang}, \citenamefont {Wang},\ and\ \citenamefont
  {Ding}}]{QianIOP}%
  \BibitemOpen
  \bibfield  {author} {\bibinfo {author} {\bibfnamefont {T.}~\bibnamefont
  {Qian}}, \bibinfo {author} {\bibfnamefont {H.}~\bibnamefont {Miao}}, \bibinfo
  {author} {\bibfnamefont {Z.~J.}\ \bibnamefont {Wang}}, \bibinfo {author}
  {\bibfnamefont {X.}~\bibnamefont {Shi}}, \bibinfo {author} {\bibfnamefont
  {Y.~B.}\ \bibnamefont {Huang}}, \bibinfo {author} {\bibfnamefont
  {P.}~\bibnamefont {Zhang}}, \bibinfo {author} {\bibfnamefont
  {N.}~\bibnamefont {Xu}}, \bibinfo {author} {\bibfnamefont {L.~K.}\
  \bibnamefont {Zeng}}, \bibinfo {author} {\bibfnamefont {J.~Z.}\ \bibnamefont
  {Ma}}, \bibinfo {author} {\bibfnamefont {P.}~\bibnamefont {Richard}},
  \bibinfo {author} {\bibfnamefont {M.}~\bibnamefont {Shi}}, \bibinfo {author}
  {\bibfnamefont {G.}~\bibnamefont {Xu}}, \bibinfo {author} {\bibfnamefont
  {X.}~\bibnamefont {Dai}}, \bibinfo {author} {\bibfnamefont {Z.}~\bibnamefont
  {Fang}}, \bibinfo {author} {\bibfnamefont {A.~F.}\ \bibnamefont {Fang}},
  \bibinfo {author} {\bibfnamefont {N.~L.}\ \bibnamefont {Wang}}, \ and\
  \bibinfo {author} {\bibfnamefont {H.}~\bibnamefont {Ding}},\ }\href {\doibase
  10.1088/1367-2630/16/12/123038} {\bibfield  {journal} {\bibinfo  {journal}
  {New Journal of Physics}\ }\textbf {\bibinfo {volume} {16}},\ \bibinfo
  {pages} {123038} (\bibinfo {year} {2014})}\BibitemShut {NoStop}%
\bibitem [{\citenamefont {Saleh}\ and\ \citenamefont
  {Artyukhin}(2020)}]{SalehEntropy}%
  \BibitemOpen
  \bibfield  {author} {\bibinfo {author} {\bibfnamefont {G.}~\bibnamefont
  {Saleh}}\ and\ \bibinfo {author} {\bibfnamefont {S.}~\bibnamefont
  {Artyukhin}},\ }\href {\doibase 10.1021/acs.jpclett.0c00012} {\bibfield
  {journal} {\bibinfo  {journal} {The Journal of Physical Chemistry Letters}\
  }\textbf {\bibinfo {volume} {11}},\ \bibinfo {pages} {2127} (\bibinfo {year}
  {2020})},\ \bibinfo {note} {pMID: 32079398}\BibitemShut {NoStop}%
\bibitem [{\citenamefont {Muntwiler}\ \emph {et~al.}(2017)\citenamefont
  {Muntwiler}, \citenamefont {Zhang}, \citenamefont {Stania}, \citenamefont
  {Matsui}, \citenamefont {Oberta}, \citenamefont {Flechsig}, \citenamefont
  {Patthey}, \citenamefont {Quitmann}, \citenamefont {Glatzel}, \citenamefont
  {Widmer}, \citenamefont {Meyer}, \citenamefont {Jung}, \citenamefont {Aebi},
  \citenamefont {Fasel},\ and\ \citenamefont {Greber}}]{PEARL}%
  \BibitemOpen
  \bibfield  {author} {\bibinfo {author} {\bibfnamefont {M.}~\bibnamefont
  {Muntwiler}}, \bibinfo {author} {\bibfnamefont {J.}~\bibnamefont {Zhang}},
  \bibinfo {author} {\bibfnamefont {R.}~\bibnamefont {Stania}}, \bibinfo
  {author} {\bibfnamefont {F.}~\bibnamefont {Matsui}}, \bibinfo {author}
  {\bibfnamefont {P.}~\bibnamefont {Oberta}}, \bibinfo {author} {\bibfnamefont
  {U.}~\bibnamefont {Flechsig}}, \bibinfo {author} {\bibfnamefont
  {L.}~\bibnamefont {Patthey}}, \bibinfo {author} {\bibfnamefont
  {C.}~\bibnamefont {Quitmann}}, \bibinfo {author} {\bibfnamefont
  {T.}~\bibnamefont {Glatzel}}, \bibinfo {author} {\bibfnamefont
  {R.}~\bibnamefont {Widmer}}, \bibinfo {author} {\bibfnamefont
  {E.}~\bibnamefont {Meyer}}, \bibinfo {author} {\bibfnamefont
  {T.}~\bibnamefont {Jung}}, \bibinfo {author} {\bibfnamefont {P.}~\bibnamefont
  {Aebi}}, \bibinfo {author} {\bibfnamefont {R.}~\bibnamefont {Fasel}}, \ and\
  \bibinfo {author} {\bibfnamefont {T.}~\bibnamefont {Greber}},\ }\href
  {\doibase 10.1107/S1600577516018646} {\bibfield  {journal} {\bibinfo
  {journal} {Journal of Synchrotron Radiation}\ }\textbf {\bibinfo {volume}
  {24}},\ \bibinfo {pages} {354} (\bibinfo {year} {2017})}\BibitemShut
  {NoStop}%
\bibitem [{\citenamefont {Kresse}\ and\ \citenamefont
  {Hafner}(1993)}]{kresse1993}%
  \BibitemOpen
  \bibfield  {author} {\bibinfo {author} {\bibfnamefont {G.}~\bibnamefont
  {Kresse}}\ and\ \bibinfo {author} {\bibfnamefont {J.}~\bibnamefont
  {Hafner}},\ }\href {\doibase 10.1103/PhysRevB.47.558} {\bibfield  {journal}
  {\bibinfo  {journal} {Phys. Rev. B}\ }\textbf {\bibinfo {volume} {47}},\
  \bibinfo {pages} {558} (\bibinfo {year} {1993})}\BibitemShut {NoStop}%
\bibitem [{\citenamefont {Kresse}\ and\ \citenamefont
  {Hafner}(1994)}]{kresse1994}%
  \BibitemOpen
  \bibfield  {author} {\bibinfo {author} {\bibfnamefont {G.}~\bibnamefont
  {Kresse}}\ and\ \bibinfo {author} {\bibfnamefont {J.}~\bibnamefont
  {Hafner}},\ }\href {\doibase 10.1103/PhysRevB.49.14251} {\bibfield  {journal}
  {\bibinfo  {journal} {Phys. Rev. B}\ }\textbf {\bibinfo {volume} {49}},\
  \bibinfo {pages} {14251} (\bibinfo {year} {1994})}\BibitemShut {NoStop}%
\bibitem [{\citenamefont {Kresse}\ and\ \citenamefont
  {Furthmueller}(1996{\natexlab{a}})}]{kresse1996a}%
  \BibitemOpen
  \bibfield  {author} {\bibinfo {author} {\bibfnamefont {G.}~\bibnamefont
  {Kresse}}\ and\ \bibinfo {author} {\bibfnamefont {J.}~\bibnamefont
  {Furthmueller}},\ }\href {\doibase
  https://doi.org/10.1016/0927-0256(96)00008-0} {\bibfield  {journal} {\bibinfo
   {journal} {Computational Materials Science}\ }\textbf {\bibinfo {volume}
  {6}},\ \bibinfo {pages} {15 } (\bibinfo {year}
  {1996}{\natexlab{a}})}\BibitemShut {NoStop}%
\bibitem [{\citenamefont {Kresse}\ and\ \citenamefont
  {Furthmueller}(1996{\natexlab{b}})}]{kresse1996b}%
  \BibitemOpen
  \bibfield  {author} {\bibinfo {author} {\bibfnamefont {G.}~\bibnamefont
  {Kresse}}\ and\ \bibinfo {author} {\bibfnamefont {J.}~\bibnamefont
  {Furthmueller}},\ }\href {\doibase 10.1103/PhysRevB.54.11169} {\bibfield
  {journal} {\bibinfo  {journal} {Phys. Rev. B}\ }\textbf {\bibinfo {volume}
  {54}},\ \bibinfo {pages} {11169} (\bibinfo {year}
  {1996}{\natexlab{b}})}\BibitemShut {NoStop}%
\bibitem [{\citenamefont {Kresse}\ and\ \citenamefont
  {Joubert}(1999)}]{kresse1999}%
  \BibitemOpen
  \bibfield  {author} {\bibinfo {author} {\bibfnamefont {G.}~\bibnamefont
  {Kresse}}\ and\ \bibinfo {author} {\bibfnamefont {D.}~\bibnamefont
  {Joubert}},\ }\href {\doibase 10.1103/PhysRevB.59.1758} {\bibfield  {journal}
  {\bibinfo  {journal} {Phys. Rev. B}\ }\textbf {\bibinfo {volume} {59}},\
  \bibinfo {pages} {1758} (\bibinfo {year} {1999})}\BibitemShut {NoStop}%
\bibitem [{\citenamefont {Perdew}\ \emph {et~al.}(1996)\citenamefont {Perdew},
  \citenamefont {Burke},\ and\ \citenamefont {Ernzerhof}}]{perdew1996}%
  \BibitemOpen
  \bibfield  {author} {\bibinfo {author} {\bibfnamefont {J.~P.}\ \bibnamefont
  {Perdew}}, \bibinfo {author} {\bibfnamefont {K.}~\bibnamefont {Burke}}, \
  and\ \bibinfo {author} {\bibfnamefont {M.}~\bibnamefont {Ernzerhof}},\ }\href
  {\doibase 10.1103/PhysRevLett.77.3865} {\bibfield  {journal} {\bibinfo
  {journal} {Phys. Rev. Lett.}\ }\textbf {\bibinfo {volume} {77}},\ \bibinfo
  {pages} {3865} (\bibinfo {year} {1996})}\BibitemShut {NoStop}%
\bibitem [{\citenamefont {Medeiros}\ \emph {et~al.}(2014)\citenamefont
  {Medeiros}, \citenamefont {Stafstroem},\ and\ \citenamefont
  {Bjoerk}}]{medeiros2014}%
  \BibitemOpen
  \bibfield  {author} {\bibinfo {author} {\bibfnamefont {P.~V.~C.}\
  \bibnamefont {Medeiros}}, \bibinfo {author} {\bibfnamefont {S.}~\bibnamefont
  {Stafstroem}}, \ and\ \bibinfo {author} {\bibfnamefont {J.}~\bibnamefont
  {Bjoerk}},\ }\href {\doibase 10.1103/PhysRevB.89.041407} {\bibfield
  {journal} {\bibinfo  {journal} {Phys. Rev. B}\ }\textbf {\bibinfo {volume}
  {89}},\ \bibinfo {pages} {041407} (\bibinfo {year} {2014})}\BibitemShut
  {NoStop}%
\bibitem [{\citenamefont {Medeiros}\ \emph {et~al.}(2015)\citenamefont
  {Medeiros}, \citenamefont {Tsirkin}, \citenamefont {Stafstroem},\ and\
  \citenamefont {Bjoerk}}]{medeiros2015}%
  \BibitemOpen
  \bibfield  {author} {\bibinfo {author} {\bibfnamefont {P.~V.~C.}\
  \bibnamefont {Medeiros}}, \bibinfo {author} {\bibfnamefont {S.~S.}\
  \bibnamefont {Tsirkin}}, \bibinfo {author} {\bibfnamefont {S.}~\bibnamefont
  {Stafstroem}}, \ and\ \bibinfo {author} {\bibfnamefont {J.}~\bibnamefont
  {Bjoerk}},\ }\href {\doibase 10.1103/PhysRevB.91.041116} {\bibfield
  {journal} {\bibinfo  {journal} {Phys. Rev. B}\ }\textbf {\bibinfo {volume}
  {91}},\ \bibinfo {pages} {041116} (\bibinfo {year} {2015})}\BibitemShut
  {NoStop}%
\bibitem [{\citenamefont {Ootsuki}\ \emph {et~al.}(2013)\citenamefont
  {Ootsuki}, \citenamefont {Pyon}, \citenamefont {Kudo}, \citenamefont
  {Nohara}, \citenamefont {Horio}, \citenamefont {Yoshida}, \citenamefont
  {Fujimori}, \citenamefont {Arita}, \citenamefont {Anzai}, \citenamefont
  {Namatame}, \citenamefont {Taniguchi}, \citenamefont {L.~Saini},\ and\
  \citenamefont {Mizokawa}}]{OotsukiJapanLettersES}%
  \BibitemOpen
  \bibfield  {author} {\bibinfo {author} {\bibfnamefont {D.}~\bibnamefont
  {Ootsuki}}, \bibinfo {author} {\bibfnamefont {S.}~\bibnamefont {Pyon}},
  \bibinfo {author} {\bibfnamefont {K.}~\bibnamefont {Kudo}}, \bibinfo {author}
  {\bibfnamefont {M.}~\bibnamefont {Nohara}}, \bibinfo {author} {\bibfnamefont
  {M.}~\bibnamefont {Horio}}, \bibinfo {author} {\bibfnamefont
  {T.}~\bibnamefont {Yoshida}}, \bibinfo {author} {\bibfnamefont
  {A.}~\bibnamefont {Fujimori}}, \bibinfo {author} {\bibfnamefont
  {M.}~\bibnamefont {Arita}}, \bibinfo {author} {\bibfnamefont
  {H.}~\bibnamefont {Anzai}}, \bibinfo {author} {\bibfnamefont
  {H.}~\bibnamefont {Namatame}}, \bibinfo {author} {\bibfnamefont
  {M.}~\bibnamefont {Taniguchi}}, \bibinfo {author} {\bibfnamefont
  {N.}~\bibnamefont {L.~Saini}}, \ and\ \bibinfo {author} {\bibfnamefont
  {T.}~\bibnamefont {Mizokawa}},\ }\href {\doibase 10.7566/JPSJ.82.093704}
  {\bibfield  {journal} {\bibinfo  {journal} {Journal of the Physical Society
  of Japan}\ }\textbf {\bibinfo {volume} {82}},\ \bibinfo {pages} {093704}
  (\bibinfo {year} {2013})}\BibitemShut {NoStop}%
\bibitem [{\citenamefont {Ootsuki}\ \emph {et~al.}(2014)\citenamefont
  {Ootsuki}, \citenamefont {Toriyama}, \citenamefont {Kobayashi}, \citenamefont
  {Pyon}, \citenamefont {Kudo}, \citenamefont {Nohara}, \citenamefont
  {Sugimoto}, \citenamefont {Yoshida}, \citenamefont {Horio}, \citenamefont
  {Fujimori}, \citenamefont {Arita}, \citenamefont {Anzai}, \citenamefont
  {Namatame}, \citenamefont {Taniguchi}, \citenamefont {Saini}, \citenamefont
  {Konishi}, \citenamefont {Ohta},\ and\ \citenamefont
  {Mizokawa}}]{OotsukiJapanLettersPt}%
  \BibitemOpen
  \bibfield  {author} {\bibinfo {author} {\bibfnamefont {D.}~\bibnamefont
  {Ootsuki}}, \bibinfo {author} {\bibfnamefont {T.}~\bibnamefont {Toriyama}},
  \bibinfo {author} {\bibfnamefont {M.}~\bibnamefont {Kobayashi}}, \bibinfo
  {author} {\bibfnamefont {S.}~\bibnamefont {Pyon}}, \bibinfo {author}
  {\bibfnamefont {K.}~\bibnamefont {Kudo}}, \bibinfo {author} {\bibfnamefont
  {M.}~\bibnamefont {Nohara}}, \bibinfo {author} {\bibfnamefont
  {T.}~\bibnamefont {Sugimoto}}, \bibinfo {author} {\bibfnamefont
  {T.}~\bibnamefont {Yoshida}}, \bibinfo {author} {\bibfnamefont
  {M.}~\bibnamefont {Horio}}, \bibinfo {author} {\bibfnamefont
  {A.}~\bibnamefont {Fujimori}}, \bibinfo {author} {\bibfnamefont
  {M.}~\bibnamefont {Arita}}, \bibinfo {author} {\bibfnamefont
  {H.}~\bibnamefont {Anzai}}, \bibinfo {author} {\bibfnamefont
  {H.}~\bibnamefont {Namatame}}, \bibinfo {author} {\bibfnamefont
  {M.}~\bibnamefont {Taniguchi}}, \bibinfo {author} {\bibfnamefont {N.~L.}\
  \bibnamefont {Saini}}, \bibinfo {author} {\bibfnamefont {T.}~\bibnamefont
  {Konishi}}, \bibinfo {author} {\bibfnamefont {Y.}~\bibnamefont {Ohta}}, \
  and\ \bibinfo {author} {\bibfnamefont {T.}~\bibnamefont {Mizokawa}},\ }\href
  {\doibase 10.7566/JPSJ.83.033704} {\bibfield  {journal} {\bibinfo  {journal}
  {Journal of the Physical Society of Japan}\ }\textbf {\bibinfo {volume}
  {83}},\ \bibinfo {pages} {033704} (\bibinfo {year} {2014})}\BibitemShut
  {NoStop}%
\bibitem [{\citenamefont {Ootsuki}\ \emph {et~al.}(2019)\citenamefont
  {Ootsuki}, \citenamefont {Ishii}, \citenamefont {Kudo}, \citenamefont
  {Nohara}, \citenamefont {Arita}, \citenamefont {Namatame}, \citenamefont
  {Taniguchi}, \citenamefont {Saini},\ and\ \citenamefont
  {Mizokawa}}]{OotsukiJournPhys}%
  \BibitemOpen
  \bibfield  {author} {\bibinfo {author} {\bibfnamefont {D.}~\bibnamefont
  {Ootsuki}}, \bibinfo {author} {\bibfnamefont {H.}~\bibnamefont {Ishii}},
  \bibinfo {author} {\bibfnamefont {K.}~\bibnamefont {Kudo}}, \bibinfo {author}
  {\bibfnamefont {M.}~\bibnamefont {Nohara}}, \bibinfo {author} {\bibfnamefont
  {M.}~\bibnamefont {Arita}}, \bibinfo {author} {\bibfnamefont
  {H.}~\bibnamefont {Namatame}}, \bibinfo {author} {\bibfnamefont
  {M.}~\bibnamefont {Taniguchi}}, \bibinfo {author} {\bibfnamefont {N.~L.}\
  \bibnamefont {Saini}}, \ and\ \bibinfo {author} {\bibfnamefont
  {T.}~\bibnamefont {Mizokawa}},\ }\href {\doibase
  https://doi.org/10.1016/j.jpcs.2018.02.015} {\bibfield  {journal} {\bibinfo
  {journal} {Journal of Physics and Chemistry of Solids}\ }\textbf {\bibinfo
  {volume} {128}},\ \bibinfo {pages} {270 } (\bibinfo {year}
  {2019})}\BibitemShut {NoStop}%
\bibitem [{\citenamefont {Lee}\ \emph {et~al.}(2017)\citenamefont {Lee},
  \citenamefont {Ko}, \citenamefont {Kim}, \citenamefont {Park}, \citenamefont
  {Yang}, \citenamefont {Cheong},\ and\ \citenamefont {Park}}]{LeeIOP}%
  \BibitemOpen
  \bibfield  {author} {\bibinfo {author} {\bibfnamefont {H.}~\bibnamefont
  {Lee}}, \bibinfo {author} {\bibfnamefont {K.-T.}\ \bibnamefont {Ko}},
  \bibinfo {author} {\bibfnamefont {K.}~\bibnamefont {Kim}}, \bibinfo {author}
  {\bibfnamefont {B.-G.}\ \bibnamefont {Park}}, \bibinfo {author}
  {\bibfnamefont {J.}~\bibnamefont {Yang}}, \bibinfo {author} {\bibfnamefont
  {S.-W.}\ \bibnamefont {Cheong}}, \ and\ \bibinfo {author} {\bibfnamefont
  {J.-H.}\ \bibnamefont {Park}},\ }\href {\doibase 10.1209/0295-5075/120/47003}
  {\bibfield  {journal} {\bibinfo  {journal} {Europhysics Letters}\ }\textbf
  {\bibinfo {volume} {120}},\ \bibinfo {pages} {47003} (\bibinfo {year}
  {2017})}\BibitemShut {NoStop}%
\bibitem [{\citenamefont {Bahramy}\ \emph {et~al.}(2017)\citenamefont
  {Bahramy}, \citenamefont {Clark}, \citenamefont {Yang}, \citenamefont {Feng},
  \citenamefont {Bawden}, \citenamefont {Riley}, \citenamefont {Markovic},
  \citenamefont {Mazzola}, \citenamefont {Sunko}, \citenamefont {Biswas},
  \citenamefont {Cooil}, \citenamefont {Jorge}, \citenamefont {Wells},
  \citenamefont {Leandersson}, \citenamefont {Balasubramanian}, \citenamefont
  {Fujii}, \citenamefont {Vobornik}, \citenamefont {Rault}, \citenamefont
  {Kim}, \citenamefont {Hoesch}, \citenamefont {Okawa}, \citenamefont {Askawa},
  \citenamefont {Sasagawa}, \citenamefont {Eknapakul}, \citenamefont
  {Meevasana},\ and\ \citenamefont {King}}]{KingNatMat}%
  \BibitemOpen
  \bibfield  {author} {\bibinfo {author} {\bibfnamefont {M.~S.}\ \bibnamefont
  {Bahramy}}, \bibinfo {author} {\bibfnamefont {O.~J.}\ \bibnamefont {Clark}},
  \bibinfo {author} {\bibfnamefont {B.-J.}\ \bibnamefont {Yang}}, \bibinfo
  {author} {\bibfnamefont {J.}~\bibnamefont {Feng}}, \bibinfo {author}
  {\bibfnamefont {L.}~\bibnamefont {Bawden}}, \bibinfo {author} {\bibfnamefont
  {J.~M.}\ \bibnamefont {Riley}}, \bibinfo {author} {\bibfnamefont
  {I.}~\bibnamefont {Markovic}}, \bibinfo {author} {\bibfnamefont
  {F.}~\bibnamefont {Mazzola}}, \bibinfo {author} {\bibfnamefont
  {V.}~\bibnamefont {Sunko}}, \bibinfo {author} {\bibfnamefont
  {D.}~\bibnamefont {Biswas}}, \bibinfo {author} {\bibfnamefont {S.~P.}\
  \bibnamefont {Cooil}}, \bibinfo {author} {\bibfnamefont {M.}~\bibnamefont
  {Jorge}}, \bibinfo {author} {\bibfnamefont {J.~W.}\ \bibnamefont {Wells}},
  \bibinfo {author} {\bibfnamefont {M.}~\bibnamefont {Leandersson}}, \bibinfo
  {author} {\bibfnamefont {T.}~\bibnamefont {Balasubramanian}}, \bibinfo
  {author} {\bibfnamefont {J.}~\bibnamefont {Fujii}}, \bibinfo {author}
  {\bibfnamefont {I.}~\bibnamefont {Vobornik}}, \bibinfo {author}
  {\bibfnamefont {J.~E.}\ \bibnamefont {Rault}}, \bibinfo {author}
  {\bibfnamefont {T.~K.}\ \bibnamefont {Kim}}, \bibinfo {author} {\bibfnamefont
  {M.}~\bibnamefont {Hoesch}}, \bibinfo {author} {\bibfnamefont
  {K.}~\bibnamefont {Okawa}}, \bibinfo {author} {\bibfnamefont
  {M.}~\bibnamefont {Askawa}}, \bibinfo {author} {\bibfnamefont
  {T.}~\bibnamefont {Sasagawa}}, \bibinfo {author} {\bibfnamefont
  {T.}~\bibnamefont {Eknapakul}}, \bibinfo {author} {\bibfnamefont
  {W.}~\bibnamefont {Meevasana}}, \ and\ \bibinfo {author} {\bibfnamefont
  {P.~D.~C.}\ \bibnamefont {King}},\ }\href {https://doi.org/10.1038/nmat5031}
  {\bibfield  {journal} {\bibinfo  {journal} {Nature Materials}\ }\textbf
  {\bibinfo {volume} {17}},\ \bibinfo {pages} {21} (\bibinfo {year}
  {2017})}\BibitemShut {NoStop}%
\bibitem [{Note1()}]{Note1}%
  \BibitemOpen
  \bibinfo {note} {Formally, we do not measure electronic surface states with
  $\mu $LEED, but we make use of the transition temperatures (upon warming)
  extracted with $\mu $LEED to extrapolate the phase diagram of Fig.~\ref
  {Figure UV}~(f).}\BibitemShut {Stop}%
\bibitem [{\citenamefont {Ivashko}\ \emph {et~al.}(2017)\citenamefont
  {Ivashko}, \citenamefont {Yang}, \citenamefont {Destraz}, \citenamefont
  {Martino}, \citenamefont {Chen}, \citenamefont {Guo}, \citenamefont {Yuan},
  \citenamefont {Pisoni}, \citenamefont {Matus}, \citenamefont {Pyon},
  \citenamefont {Kudo}, \citenamefont {Nohara}, \citenamefont {Forr\`{o}},
  \citenamefont {Ronnow}, \citenamefont {Huecker}, \citenamefont {Zimmermann},\
  and\ \citenamefont {Chang}}]{Ivashko2017}%
  \BibitemOpen
  \bibfield  {author} {\bibinfo {author} {\bibfnamefont {O.}~\bibnamefont
  {Ivashko}}, \bibinfo {author} {\bibfnamefont {L.}~\bibnamefont {Yang}},
  \bibinfo {author} {\bibfnamefont {D.}~\bibnamefont {Destraz}}, \bibinfo
  {author} {\bibfnamefont {E.}~\bibnamefont {Martino}}, \bibinfo {author}
  {\bibfnamefont {Y.}~\bibnamefont {Chen}}, \bibinfo {author} {\bibfnamefont
  {C.}~\bibnamefont {Guo}}, \bibinfo {author} {\bibfnamefont {H.}~\bibnamefont
  {Yuan}}, \bibinfo {author} {\bibfnamefont {A.}~\bibnamefont {Pisoni}},
  \bibinfo {author} {\bibfnamefont {P.}~\bibnamefont {Matus}}, \bibinfo
  {author} {\bibfnamefont {S.}~\bibnamefont {Pyon}}, \bibinfo {author}
  {\bibfnamefont {K.}~\bibnamefont {Kudo}}, \bibinfo {author} {\bibfnamefont
  {M.}~\bibnamefont {Nohara}}, \bibinfo {author} {\bibfnamefont
  {L.}~\bibnamefont {Forr\`{o}}}, \bibinfo {author} {\bibfnamefont
  {H.}~\bibnamefont {Ronnow}}, \bibinfo {author} {\bibfnamefont
  {M.}~\bibnamefont {Huecker}}, \bibinfo {author} {\bibfnamefont {M.~v.}\
  \bibnamefont {Zimmermann}}, \ and\ \bibinfo {author} {\bibfnamefont
  {J.}~\bibnamefont {Chang}},\ }\href {\doibase 0.1038/s41598-017-16945-7}
  {\bibfield  {journal} {\bibinfo  {journal} {Scientific Reports}\ }\textbf
  {\bibinfo {volume} {7}},\ \bibinfo {pages} {17157} (\bibinfo {year}
  {2017})}\BibitemShut {NoStop}%
\end{thebibliography}

%

\end{document}